\documentclass[12pt,preprint]{aastex}

\begin{document}

\title{A Deep {\em Hubble Space Telescope} H-Band Imaging Survey of
  Massive Gas-Rich Mergers. II. The QUEST QSOs \footnote{Based on
  observations with the NASA/ESA {\em Hubble Space Telescope},
  obtained at the Space Telescope Science Institute, which is operated
  by the Association of Universities for Research in Astronomy,
  Inc. under NASA contract No.  NAS5-26555.}}

\author{S. Veilleux\footnote{Also: Max-Planck-Institut f\"ur
    extraterrestrische Physik, Postfach 1312, D-85741 Garching,
    Germany}, D.-C. Kim\footnote{New address: Department of
Astronomy, University of Virginia, Charlottesville, VA 22904-4325},
and D. S. N. Rupke\footnote{New address: Institute for Astronomy,
University of Hawaii, 2680 Woodlawn Drive, Honolulu, HI 96822.}}
\affil{Department of Astronomy, University of Maryland, College Park,
MD 20742; \\ veilleux@astro.umd.edu, ddk3wc@mail.astro.virginia.edu,
drupke@ifa.hawaii.edu}

\author{C. Y. Peng} \affil{Herzberg Institute of Astrophysics,
National Research Council of Canada, 5071 West Saanich Rd., Victoria,
BC V9E 2E7, Canada; cyp@nrc-cnrc.gc.ca}

\author{L. J. Tacconi, R. Genzel, D. Lutz, E. Sturm, A. Contursi, and
M. Schweitzer} \affil{Max-Planck-Institut f\"ur extraterrestrische
Physik, Postfach 1312, D-85741 Garching, Germany; linda@mpe.mpg.de,
genzel@mpe.mpg.de, lutz@mpe.mpg.de, sturm@mpe.mpg.de,
contursi@mpe.mpg.de, schweitzer@mpe.mpg.de}

\author{K. M. Dasyra} \affil{Spitzer Science Center, Mail Code 220-6,
  California Institute of Technology, 1200 East California Boulevard,
  Pasadena, CA 91125; dasyra@ipac.caltech.edu}

\author{L. C. Ho} \affil{The Observatories of the Carnegie Institution
  of Washington, 813 Santa Barbara St., Pasadena, CA 91101;
  lho@ociw.edu}

\author{D. B. Sanders}
\affil{Institute for Astronomy, University of Hawaii, 2680 Woodlawn
  Drive, Honolulu, HI 96822; sanders@ifa.hawaii.edu}

\author{A. Burkert} \affil{University Observatory Munich,
Scheinerstrasse 1, D-81679 Munich, Germany; burkert@usm.uni-muenchen.de}

\begin{abstract}
  We report the results from a deep {\em HST} NICMOS H-band imaging
  survey of 28 $z < 0.3$ QSOs from the Palomar-Green (PG) sample. This
  program is part of {\em QUEST} ({\em Q}uasar / {\em U}LIRG {\em
    E}volution {\em ST}udy) and complements a similar set of data on
  26 highly-nucleated ULIRGs presented in Paper I.  Our analysis
  indicates that the fraction of QSOs with elliptical hosts is higher
  among QSOs with undetected far-infrared (FIR) emission, small
  infrared excess ($L_{\rm IR}/L_{\rm B} < 10$), and luminous hosts.
  The hosts of FIR-faint QSOs show a tendency to have less pronounced
  merger-induced morphological anomalies and larger QSO-to-host
  luminosity ratios on average than the hosts of FIR-bright QSOs,
  consistent with late-merger evolution from FIR-bright to FIR-faint
  QSOs.  The spheroid sizes ($\sim$0.3 -- 5.5 kpc) and total host
  luminosities ($\sim$0.6 -- 7.2 L$^*_H$) of the radio-quiet PG~QSOs
  in our sample are statistically indistinguishable from the ULIRG
  hosts presented in Paper I, while those of radio-loud PG~QSOs are
  systematically larger and more luminous.  ULIRGs and PG~QSOs with
  elliptical hosts fall near, but not exactly on, the fundamental
  plane of inactive spheroids. We confirm the systematic trend noted
  in Paper I for objects with small ($\la$ 2 kpc) spheroids to be up
  to $\sim$ 1 mag.\ brighter than inactive spheroids.  The host colors
  and wavelength dependence of their sizes support the idea that these
  deviations are due at least in part to non-nuclear star
  formation. However, the amplitudes of these deviations depend mainly
  on host sizes, and possibly on infrared excess, but not on merger
  phase, QSO-to-host luminosity ratio, optical spectral type, AGN
  fractional contribution to the bolometric luminosity, or host R$-$H
  color.  Taken at face value (i.e., no correction for extinction or
  the presence of a young stellar population), the H-band
  spheroid-host luminosities imply black hole masses $\sim$ 5 -- 200
  $\times$ 10$^7$ M$_\odot$ and sub-Eddington mass accretion rates for
  both QSOs and ULIRGs. These results are compared with published
  black hole mass estimates derived from other methods.
\end{abstract}


\section{Introduction}

In this series of papers, the structural properties of massive
gas-rich mergers in the local universe are derived to provide insights
into galaxy merging, a key driving force of galaxy evolution over the
history of the universe.  This study is part of a comprehensive
program called QUEST - Quasar/ULIRG Evolution Study.  This program
combines optical and near-infrared imaging and spectroscopic data
obtained from the ground with H-band imaging and mid-infrared
spectroscopic data obtained with the {\em Hubble Space Telescope}
({\em HST}) and {\em Spitzer Space Telescope}, respectively (see
Veilleux et al.\ 2009 and references therein, for a more detailed
discussion of QUEST).

In Veilleux et al.\ (2006; hereafter Paper I), we reported the results
from a Cycle 12 {\em HST} NICMOS H-band imaging survey of 26 $z < 0.3$
ULIRGs and 7 infrared-bright Palomar-Green (PG) QSOs.  Unsuspected
double nuclei were detected in 5 ULIRGs.  The great majority (81\%) of
the single-nucleus systems showed a prominent elliptical-like
morphology.  However, low-surface-brightness exponential disks were
detected on large scale in at least 4 of these sources.  The hosts of
`warm' ({\em IRAS} 25-to-60 $\mu$m flux ratio, $f_{25}/f_{60} > 0.2$),
AGN-like systems were found to be elliptical-like and have less
pronounced merger-induced morphological anomalies than the hosts of
cool systems with LINER or HII region-like nuclear optical spectral
types, suggestive of merger-driven evolution from cool to warm ULIRGs.
The host sizes and luminosities of the 7 PG~QSOs in the sample were
statistically indistinguishable from those of the ULIRG hosts.  The
hosts of ULIRGs and PG~QSOs were found to lie close to the locations
of intermediate-luminosity ($\sim$ 0.5 -- 2 $L^*$) spheroids in the
photometric projection of the fundamental plane (FP) of
ellipticals. However, ULIRGs with small hosts were found to be
generally brighter than normal inactive spheroids, possibly due to
excess near-infrared emission from a circumnuclear starburst.

The 7 PG~QSOs in the Cycle 12 sample were selected to be far-infrared
(FIR) brighter than typical PG~QSOs. Netzer et al.\ (2007) have
recently argued that the bulk of the infrared luminosity, $L_{\rm IR}
\equiv L(8 - 1000 \mu m)$, in PG~QSOs is produced by a massive
starburst. So it is not clear whether the Cycle 12 results can be
generalized to PG~QSOs as a whole.  Here we report the results from an
analysis of 21 additional PG~QSOs to address this issue.  In Section
2, we describe the sample of QSOs used in our study along the
extensive ancillary dataset on these objects.  Our methods to obtain,
reduce, and analyze the data are very similar to those used in Paper
I, so we discuss them only briefly in Section 3. The results are
presented in Section 4 and discussed and compared with those of
earlier studies in Section 5. The main conclusions are summarized in
Section 6.  Throughout this paper, we adopt $H_0$ = 75 km s$^{-1}$
Mpc$^{-1}$, $\Omega_M$ = 0.3, and $\Omega_\Lambda$ = 0.7.

\section{QSO Sample \& Ancillary Data}

The QSO component of QUEST has already been discussed in detail in
Schweitzer et al.\ (2006, 2008), Netzer et al.\ (2007), and Veilleux
et al.\ (2009) and this detailed discussion will not be repeated here.
The Cycle 15 {\em HST} sample consists of 23 $z \la 0.3$ quasars,
including 22 Palomar-Green (PG) quasars from the Bright Quasar Sample
(Schmidt \& Green 1983) and another one (B2~2201+31A = 4C 31.63) with
a $B$ magnitude that actually satisfies the PG QSO completeness
criterion of Schmidt \& Green (1983). Failure to acquire guide stars
with the Fine Guidance Sensors severely degraded the observations for
two of these objects (PG~0953+414 and PG~1004+130), so they are not
discussed any further in the paper.

The combined sample of 28 PG~QSOs successfully observed during Cycles
12 and 15 covers the low-redshift and low B-band luminosity ends of
the PG~QSO sample (see Figure 2 in Veilleux et al.\ 2009), and they are
well matched in redshift with the 1-Jy ULIRGs studied in Paper I
(Figure 3 of Veilleux et al.\ 2009). The combined sample of QSOs is
representative of the entire PG~QSO sample in terms of infrared excess
(i.e.,   the infrared-to-blue luminosity ratio, $L_{IR}/L_{B}$),
and FIR brightness [$L$(60 $\mu$m)/$L$(15 $\mu$m), Netzer et al.\
2007]. Table 1 lists some of the properties of the QSOs in our study,
including those observed during Cycle 12.  Note that three of the QSOs
in the {\em HST} sample (PG~1119+120, 1126$-$041, and 1229+204) have
absolute B-band magnitudes which are fainter than the traditional
luminosity threshold of QSOs ($M_B = -23$ for $H_0 = 50$ km s$^{-1}$
Mpc$^{-1}$ and $q_0$ = 0 or $M_B$ = -22.3 for the cosmology adopted
here).

An extensive set of data already exists on these objects.
Ground-based optical and near-infrared images of many of these objects
have been obtained in tip-tilt mode (Surace et al.\ 2001) and
adaptive-optics mode (Guyon et al.\ 2006), providing a spatial
resolution of 0$\farcs$2 -- 0$\farcs$8 and 0$\farcs$13-0$\farcs$30,
respectively, i.e.,  only slightly poorer than that of the {\em
  HST} data presented here ($\sim$ 0$\farcs$14).  However, a key
advantage of the present data over previous data sets is the stability
of the {\em HST} point-spread function, which allows us to derive
reliable structural parameters of the QSO hosts well within 1\arcsec\
of the center. Archival optical {\em HST} images of several local type
1 AGN, including some from the present sample, were recently
analyzed by two in our group (C.Y.P., L.C.H.)  and published as Kim et
al.\ (2008a), and also independently studied by Hamilton et al.\
(2008). The results from these previous studies are compared with
those from our {\em HST} survey in Sections 4 and 5.  All of these QSOs
have also been studied spectroscopically at optical wavelengths from
the ground (Boroson \& Green 1992) and in the mid-infrared with {\em
  Spitzer} (Schweizer et al.\ 2006, 2008; Netzer et al.\ 2007;
Veilleux et al.\ 2009). This last dataset provides valuable
information on the AGN contribution to the bolometric luminosities of
these objects; we make use of this information in Section 5. In addition,
VLT/Keck near-infrared spectroscopic data exist for a number of these
objects. Dynamical estimates of the masses of the hosts were derived
from these data (Dasyra et al.\ 2007) and are compared with our
photometric estimates in Section 5.

\section{Data Acquisition, Reduction, and Analysis}

The methods used to acquire, reduce and analyze the present data are
nearly identical to those of Paper I, so only a summary is given
below; interested readers should refer to Paper I for more detail. 

The main driver of our Cycle 15 program on the QSOs was to match the
observational setup (instrument, filter, detection level, dither
pattern) used for our Cycle-12 data to facilitate comparisons between
the two datasets. Our Cycle-12 results have confirmed that the
excellent spatial resolution and sensitivity of {\em HST} NICMOS in
the non-thermal infrared are required to extract the central point
sources from our targets and derive accurate structural parameters on
the hosts. NICMOS is better suited for this program than ACS to reduce
the impact of dust extinction and star formation on the measurements
(especially in the cores of ULIRGs and infrared bright QSOs) and to
exploit the contrast between QSO and elliptical hosts (e.g., McLeod \&
McLeod 2001 and references therein). The strong thermal background
makes deep observations at K unrealistic; our program therefore
focuses on the H band, roughly matching the waveband of our VLT and
Keck spectra.  The need for {\em deep} images can hardly be
overstated.  Comparisons of our Cycle-12 data with the results derived
from shallow (e.g., SNAP) HST images from the archives show that the
shallow HST data underestimate the luminosities and half-light radii
of the hosts, make profile fitting ambiguous (e.g., S\'ersic spheroid
versus exponential disk), and can even completely overlook low surface
brightness, tidal distortions or exponential disks extending
significantly beyond galactic bulges. To avoid these problems, we
tried to match the detection level (S/N $\approx$ 3) of our Cycle-12
data ($\sim$ 22.0 H mag.~arcsec$^{-2}$) by observing each target
for one full orbit (on-target exposure time of 2650 seconds).

NIC2 was selected for our Cycle 15 program, based on the requirements
of good sensitivity to low surface brightness features, excellent
spatial resolution (0$\farcs$076 pixel$^{-1}$) for accurate PSF (FWHM
= 0$\farcs$14) removal, and a field of view (19$\farcs$5 $\times$
19$\farcs$5) large enough to encompass most of the structures in our
targets. To help with the PSF subtraction, we also requested an
additional orbit to obtain a deep exposure of a star (SA 107-626) and
fully characterize the PSF at H.

Given the redshifts of our targets (z $\sim$ 0.05 -- 0.33; Table 1)
and the strengths of the emission features in ULIRGs and QSOs (see,
e.g., Veilleux et al.\ 1997, 1999; Dasyra et al.\ 2007), contamination
by emission lines (e.g., [Fe~II] $\lambda$1.644, Pa$\beta$) is at most
$\sim$ 10\% for the F160W filter, and is therefore not an issue here.
We used the logarithmic MULTIACCUM sequences to provide the largest
dynamic range and allow the calibration software to recover the bright
central point source.  The telescope was dithered between exposures to
better sample the instrumental PSF, and to aid with the recognition
and elimination of data artifacts.

The raw {\em HST} NICMOS data were first processed with the IDL
procedure {\em undopuft.pro} written by Eddie Bergeron at STScI (Space
Telescope Science Institute) to remove the electronic echoes of bright
sources and associated stripes, and subsequently processed with the
standard pipeline processing task {\em calnica} within IRAF/STSDAS to
correct for nonlinearity of the detector and removes bias value, dark
current, amplifier glow, and shading.  The IDL procedure {\em
  saa\_clean.pro} was used to remove the effects of cosmic ray
persistence (Bergeron \& Dickinson 2003).  Next, the four dithered
exposures of each object were combined using the ``drizzle'' technique
(Gonzaga et al.  1998).  For the photometric calibration of the
reduced data, a Vega-normalized magnitude for F160W (NIC2) was derived
following the recipe in the {\em HST} Data Handbook for NICMOS
(Dickinson et al.  2002) using the calibration appropriate for Cycle
15.

The two-dimensional fitting algorithm GALFIT (Peng et al.\ 2002) was
used to accurately remove the central point source in each object and
determine the structural parameters of the underlying host.  In some
cases, the analysis was carried out a second time by other members of
our group to independently verify the significance of the results.
The analysis of each object followed a number of well-defined
steps. First, we constructed a mask to exclude bright stars or small
foreground/background galaxies within the field of view. Next, we
proceeded to fit the surface brightness distribution of each object
using a single S\'ersic component (observed intensity profile $I
\propto {\rm exp}[-R^{1/n}]$) to simulate the galaxy host and a PSF
model to account for the possibility of an unresolved nuclear
starburst or AGN. The high-S/N PSF model was derived from our deep
images of SA 107-626.  The S\'ersic component was convolved with the
PSF before comparison with the data. Three S\'ersic components were
examined: $n$ = free (i.e.,   left unconstrained), $n$ = 1
(exponential disk profile), and $n$ = 4 (de Vaucouleurs profile). In
all cases, the centroids of the PSF and S\'ersic components were left
unconstrained. This relatively simple one galaxy component analysis
allowed us to get a general sense of the complexity of each system and
whether the system is disk- or spheroid-dominated.

As was the case for the ULIRGs and infrared-bright QSOs in Paper I,
the residuals from the one component galaxy fit to the QSOs are often
quite significant.  This is generally the results of merger-induced
morphological anomalies. However, in other cases, these residuals may
indicate the presence of a second low-surface-brightness galaxy
component (e.g., disk). So we decided to look into this possibility by
adding a second (PSF-convolved) galaxy component to the fits for each
object and examining the effects on the goodness of the fits. To limit
our search, we only studied the ($n$ = 1) + ($n$ = 4) case.  Here
again, the centroids of the various components were left
unconstrained. Not surprisingly given the larger number of free
parameters, these two-component models generally provide better fits
to the data.  However, a careful examination of the fitted components
often indicate that the second galaxy component is not physically
meaningful (see list of telltale signs in Section 5.2 of Paper I).  The
surest way to recognize when a second galaxy component is real is to
``put back'' into the residual image the model components individually
to see which structure was being fitted.  The components have to be
fairly distinct both spatially (e.g. axis ratio, size, centering) and
morphologically (concentration) for us to accept the two components as
being real in our assessment.

This procedure provided reliable host-galaxy structural parameters for
all but 3 objects in our sample (see Section 4.1 for a discussion of
the measurement uncertainties). For PG~1116+215, PG~1617+175, and
PG~2251+113, significant residuals due to PSF mismatches were found
near the cores of these objects.
The structural parameters derived for these objects are considered
unreliable and not included in our search for trends (Section 4) and
discussion (Section 5).

\section{Results}

\subsection{General Considerations \& Uncertainties}

The main results from the GALFIT analysis are shown in Figures 1 and 2
and listed in Tables 2 -- 5 (readers who are looking for a quick
summary of the results should refer to Table 5).  Figure 1 presents
the residuals found after subtracting one galaxy component models (PSF
+ S\'ersic with $n$ = free, 1, or 4) from the surface brightness
distributions of single-nucleus systems in our sample. In several
cases, we find that adding another S\'ersic component significantly
improves the goodness of the fits; the results of this more
sophisticated two galaxy component analysis are shown in Figure 2. The
structural parameters derived from the one and two galaxy component
fits are listed in Tables 2 and 3, respectively. Note that the exact
value of $n > 4$ is not too significant.  It generally indicates the
galaxy has either a strong core (e.g., bulge dominated) or an extended
wing (e.g. elliptical galaxies or interacting/neighboring galaxies),
or both.  Large $n$ can be caused by bad AGN subtraction, but we tried
to minimize that likelihood by using multiple components for the core.
We also tried to minimize neighboring contamination by fitting the
neighbors and/or masking.  Indeed, our images often show small
galaxies in the vicinity of the PG~QSOs, but they are considerably
fainter ($\Delta$m$_H$ $\ga$ 4 mags) than the QSO hosts.  We have no
data to determine if these small objects are associated or not with
the QSOs, so we list the magnitudes of the objects in a separate
table, Table 4, but do not discuss them any further in this paper.

Table 5 provides a summary of the best-fitting model for each object
in the sample along with a visual (hence subjective) assessment of the
presence of a stellar bar, spiral arms, and strong merger-induced
disturbances. The best-fitting models listed in this table were
adopted by inspecting the residuals in Figures 1 and 2 and the reduced
chi-squares, $\chi_\nu^2$, listed in Tables 2 and 3. The first of
these $\chi_\nu^2$ values takes into account residuals over the entire
galaxy whereas the second one excludes the central portion which is
affected by errors in the subtraction of the central PSF.  These
reduced chi-squares values should be used with caution when choosing
the best fits. First, we note that they are generally significantly
larger than unity so the fits are not formally very good. This is due
in large part to the presence of merger-induced morphological
anomalies; we return to this important point below (Section 4.3).  We also
notice that the chi-squares tend to be higher for larger, brighter,
and more PSF-dominated objects.  This is not unexpected given the
definition of $\chi_\nu^2$, which is not normalized by the intensity,
and given that the fraction of the detector area that is free of
galaxy emission is more limited for large systems than for small ones.
Thus, $\chi_\nu^2$ cannot be used to compare the goodness of fits
between objects. However, it is a useful quantity to compare the
quality of fits for the same object (the interested readers should
refer to Sections 5 and 6 of Paper I for a more detailed discussion of
the factors involved in our morphological classification).

The NIC2 observations of the QUEST sample have very high
signal-to-noise, therefore the uncertainties in the fit parameters are
generally dominated by systematic errors rather than random errors due
to Poisson noise.  Systematic errors come about from several factors,
the most common ones being a mismatch in the PSF between the data and
the model, a mismatch between the galaxy profile with the model
assumptions, or when the sky background cannot be determined
accurately for various reasons.  Even though the errors are
systematic, in AGN studies where PSF mismatch is great, there is some
randomness involved in the sense that different PSF choices we make
are drawn from a distribution around some average PSF shape.  In high
signal-to-noise, the amount of systematic error depends on the
luminosity contrast between the host galaxy and the AGN component.
The typical contrast in the QUEST sample of AGN-to-host luminosity
(Table 5, Col. 6) ranges mostly between 1 and 5, with a median of 1.5.

Kim et al. (2008b) performed very detailed AGN image fitting
simulations which can be used to estimate the uncertainty in the
fitting parameters.  Their study quantified the degree of measurement
uncertainty by drawing on different PSFs.  The scatter and systematic
errors are also presented as a function of signal-to-noise,
AGN-to-host contrast, and the size of the host galaxy, due to
different PSF choices.  Therefore we mostly draw upon that study to
infer that the systematic uncertainty for the QUEST sample to be about
10\% for the host galaxy luminosity.  The random uncertainty due to
our ignorance about the PSF are roughly: 20-50\% for the effective
radius, $\sim15$\% for the host galaxy luminosity, and $< 10$\% for
the AGN luminosity.  We can also empirically quantify the uncertainty
in the host luminosity Columns 4 (host luminosity including tidal
features) and 5 (model host luminosity) in Table 5, from which we
obtain an uncertainty of roughly $\lesssim 15$\%.

Note that the host galaxies of PG0050+124, PG0838+770, PG1229+204,
PG1426+015, and PG2214+139 cover a significant fraction of the field
of view of NIC2. The sky background is therefore difficult to
determine accurately in these images and the structural parameters of
these objects are more uncertain. This is noted in Tables 2, 3, and 5.

\subsection{Morphological Type of Host Galaxy}

The one galaxy component analysis indicates that a single spheroidal
component often provides a good fit to the surface brightness
distribution of the central portion of the PG~QSO hosts.  However, the
excellent sensitivity limit of our data allows us to also detect the
presence of faint, low-surface-brightness disks in 9/28 (32\%) objects
The results of our attempts to fit this second component as an
exponential disk are listed in Table 3 and shown in Figure 2.  Table 5
only lists the results for those nine cases where the addition of a
second, $n$ = 1 component improved the fit significantly and the
result was physically meaningful (e.g., the disk had to be concentric
with, and larger than, the bulge). Note that stellar bars are present
in at least two of these QSOs (PG~0838+770 and PG~1229+204; already
pointed out by Surace et al.\ 2001). A stellar bar may also be present
in the elliptical host of PG~1001+054, but the presence of small-scale
features in this last object limits the analysis. 

Trends are seen between morphological classification and infrared
properties.  QSOs with elliptical hosts have slightly smaller infrared
excesses (Figure 3$d$).  The median $L_{\rm IR}/L_{\rm B}$ ratio among
elliptical, bulge + disk, ambiguous hosts is 8.5, 10.5, and 10.4,
respectively. This trend fits naturally with the results of Paper I,
where we found that ULIRGs with warm 25-to-60 $\mu$m ratios, small
infrared excesses, and optical Seyfert characteristics tend to have
elliptical hosts (Figures 3$a$ and 3$b$). This trend is also consistent
with, although weaker than, that from the study of Guyon et al.\
(2006).

Interestingly, QSOs with elliptical hosts do {\em not} have
larger 25-to-60 $\mu$m ratios than those with bulge + disk or
ambiguous hosts (Figure 3$c$, median ratios of 0.32, 0.50, and 0.41,
respectively). So it appears that the trend seen in Paper I between
this ratio and the morphological classification of ULIRGs breaks down
at the smaller infrared excesses of typical QSOs.

We also note in Figure 3$e$ that all FIR-undetected QSOs have elliptical
hosts.  But this may be due to the fact that most of these QSOs are
also bolometrically luminous. Indeed, we find that the more luminous
QSOs in our sample favor elliptical hosts over late-type hosts
(Figure 3$f$).  Three of the five radio-loud QSOs in our sample have
elliptical hosts.  These results bring further support for a
luminosity and radio-loudness dependence of the host morphological
type among QSOs (e.g., Dunlop et al.\ 2003, Guyon et al.\ 2006, Paper
I; Best et al.\ 2007; Wolf \& Sheinis 2008 and references therein; see
also Section 5.1 below).

\subsection{Strength of Tidal Features}

Signs of galactic interactions such as tidal tails and bridges,
lopsided disks, distorted outer isophotes, or double nuclei are
visible in the majority (16/28 = 57\%) of the QSOs (and in all
ULIRGs, Paper I).  The residual maps in Figures 1 and 2 are a
particularly good indicator of these tidal features. Following Paper
I, we quantified the importance of these features by first adding up
the absolute values of the residuals from the best one or two galaxy
component fits over the region unaffected by the PSF subtraction and
then normalizing this quantity to the total host luminosities
(including tidal features); the results are listed in column (11) of
Tables 2 and 3. Although this quantity is sensitive to the presence of
spiral structure, dust lanes, and bright star clusters, we find in our
objects that $R_2$ is dominated by the presence of large-scale
merger-induced anomalies.

In Figure 4$a$ and 4$b$, we plot $R_2$ versus the {\em IRAS} 25-to-60
$\mu$m colors for all QSOs and ULIRGs in our sample.  PG~QSOs and warm
quasar-like ULIRGs systems tend to have smaller residuals than the
other objects in the sample. All PG~QSOs and Seyfert ULIRGs have $R_2
<$ 30\%. In Paper I, we found that ULIRGs with late-type or ambiguous
morphologies show larger residuals than elliptical systems (Figure
5$b$), suggesting that galaxies with a prominent spheroid are in the
later stages of a merger than the late-type and ambiguous systems. Our
new data on the PG~QSOs do show a similar difference between
elliptical and ambiguous systems (the residuals from the two galaxy
component fits are expected to be smaller than those from the one
galaxy component fits, so the bulge + disk systems are not considered
in our discussion).

In Figures 4$c$ and 4$d$, we compare the fit residuals with the
magnitude of the infrared excess as a function of morphological
classification and FIR strength, respectively. We find a slight trend
of increasing residuals with increasing infrared excess and FIR
strength, indicating that stronger merger-induced morphological
disturbances are found among FIR-bright QSOs with large infrared
excesses, as was suggested by Guyon et al.\ (2006).  The FIR emission
in QSOs is now believed to be primarily associated with starburst
activity (Netzer et al. 2007), so this result indicates that starburst
activity declines during the final phases of the merger process,
consistent with recent numerical simulations of major equal-mass
($\sim$ 1:1) mergers (e.g., Johansson et al.\ 2009).
Note that the presence of discernible disks in several low-luminosity
PG~QSOs can also be explained in the major merger scenario if
significant re-accretion of residual cold gas formed these disks
(e.g., Governato et al.\ 2008). In these systems, local processes such
as gas inflows along nuclear bars or spiral arms may also be
contributing to the feeding of the AGN (e.g., Storchi-Bergmann et al.\
2007 and references therein).

\subsection{Strength of Unresolved Nucleus}

Following Paper I, we quantified the importance of the PSF by
calculating the flux ratio of the PSF to the host, $I_{\rm
PSF}$/$I_{\rm host}$ using the best one or two galaxy component model
for each object.  In Paper I, we found that this ratio is less than
unity for all ULIRGs, except for all 5 ULIRGs optically classified as
Seyfert 1s. Figure 5$a$ shows that most PG~QSOs have PSF-to-host
ratios above unity, indistinguishable from those of Seyfert 1
ULIRGs. The PG~QSOs strengthen the positve correlation noted in Paper
I between the PSF-to-host ratio and {\em IRAS} 25-to-60 $\mu$m color.
The AGN therefore dominates the central H-band emission in Seyfert 1
ULIRGs and QSOs. As noted in Paper I, this result does not rule out
the possibility that a nuclear starburst is also contributing to the
PSF emission, but this starburst does not produce the bulk of the
H-band emission in the nucleus of these objects. This is consistent
with the strong dilution of the CO bandheads observed in the
near-infrared spectra of Seyfert 1 ULIRGs and PG~QSOs of Dasyra et
al.\ (2007).

A slight trend is also seen between PSF-to-host ratios and infrared
excesses (or FIR brightnesses) among QSOs: those with large infrared
excesses tend to have smaller PSF-to-host ratios (Figure 5$b$). This is
consistent with the merger scenario if FIR-bright QSOs represent an
earlier phase of QSO/merger evolution when the QSOs have not fully
emerged from their dusty cocoons.

\subsection{Host Sizes, Magnitudes, and Colors} 

Figure 6 shows the distributions of host sizes (spheroid component
only) and total (spheroid + tidal features + disk, if relevant) host
absolute magnitudes for all ULIRGs and PG~QSOs in the {\em HST}
sample.  The full range in QSO spheroid half-light radii and total
host luminosities is very broad, from $r_{1\over 2}$ = 0.3 to 9.9 kpc
and from M$_H$ = $-$23.19 to $-$26.08 or $\sim$0.6 -- 9.0 $L^*_H$,
respectively (we used M$^*_H$ = $-$23.7 mag.\ as the H-band absolute
magnitude of a $L^*$ galaxy in a Schechter function description of the
local field galaxy luminosity function; Cole et al.\ 2001; Veilleux et
al.\ 2006).  The average (median) spheroid half-light radii and total
H-band absolute magnitudes of the QSO hosts in the sample are 2.87
$\pm$ 2.59 (2.14) kpc and $-$24.60 $\pm$ 0.77 ($-$24.46) mag.  For
comparison, the same quantities for the ULIRGs in Paper I are 2.55
$\pm$ 1.43 (1.84) kpc and $-$24.06 $\pm$ 0.56 ($-$24.21) mag.  These
average QSO and ULIRG host magnitudes correspond to $\sim$ 2.3 $\pm$ 1
and $\sim$ 1.5 $\pm$ 1 L$^*_H$, respectively.

A Kologorov-Smirnov (K-S) analysis shows that the hosts of the PG~QSOs
in our sample are statistically different from the hosts of the 1-Jy
ULIRGs in terms of absolute magnitudes but not in terms of sizes
[P(null) = 2.2\% and 59\%, respectively]. A closer look at Figures
6$a$ and 6$b$ shows that the difference comes entirely from the
inclusion of radio-loud QSOs in our sample. The hosts of these systems
are systematically larger and brighter than those of the radio-quiet
QSOs in our sample ($r_{1\over 2}$ = 3.0 to 9.9 kpc and from M$_H$ =
$-$24.11 to $-$26.08 or $\sim$ 1 $-$ 9 $L^*_H$ {\em versus} $r_{1\over
  2}$ = 0.3 to 5.5 kpc and from M$_H$ = $-$23.19 to $-$25.84 or $\sim$
0.6 $-$ 7.2 $L^*_H$).  Similar differences have been found in the past
(e.g., Dunlop et al.\ 2003; Guyon et al.\ 2006; Best et al.\ 2007;
Wolf \& Sheinis 2008 and references therein).  Figure 6$b$ also shows
that QSOs with elliptical hosts display the broadest range of
luminosity, while the bulge + disk systems and the ambiguous systems
tend to populate the low- and high-luminosity ends of the
distribution, respectively.

We generally find good agreement on an object-by-object basis when
comparing our host H-band magnitudes with those of McLeod \& McLeod
(2001; two objects in common), Surace et al.\ (2001; 8 objects), and
Guyon et al.\ (2006; 20 objects). The comparisons with Surace et al.\
(2001) and Guyon et al.\ (2006) are shown in Figure 7. The Surace et
al.\ host magnitudes plotted in Figure 7$a$ were calculated by
subtracting the nuclear magnitudes from the integrated magnitudes in
their Table 2. An excellent match is found, except for one object,
PG~0007+106, which is $\sim$ 1 mag.\ brighter in the Surace et al.\
data.  This is an optically violently variable source so the
difference may be due to uncertainties in the removal of the central
PSF in the ground-based data. The Guyon et al.\ values tend to be
$\sim$ 0.4 mag. brighter than our measurements. Given the good
agreement between our data and those of Surace et al.\ and the noted
variability of the PSF in the AO data of Guyon et al., we suspect that
this shift is due to uncertainties in the PSF subtraction from these
latter data. Systematic underestimate of the background level in these
latter data could also explain the shift.

There are 13 and 10 objects in common between the present H-band study
and the archival optical {\em HST} imaging studies of Kim et al.\
(2008a) and Hamilton et al.\ (2008), respectively. The R-band (V-band)
total host magnitudes of Kim et al.\ (Hamilton et al.) are compared
with our H-band magnitudes in Figure 7$c$ (7$d$). The average (median)
R$-$H color derived from Figure 7$c$ is 1.80 $\pm$ 0.53 mag.\ (1.92
mag.).  This median value is the same as that found by Jahnke et al.\
(2004) among 19 low-redshift ($z < 0.2$) quasar host galaxies. It is
$\sim$ 0.3 mag.\ bluer than the $k$-corrected R$-$H colors of
elliptical galaxies with M$_H$ $\approx$ $-$24.5 mag.\ at $z$ $\sim$
0.2 (Lilly \& Longair 1984; Fukugita et al.\ 1995; Fioc \&
Rocca-Volmerange 1999; Jahnke et al.\ 2004; Hyv\"onen et al.\ 2007,
2008). Similarly, the median V$-$H color derived from Figure 7$d$ is
1.9 mag., considerably bluer than the $k$-corrected V$-$H colors of
elliptical galaxies with M$_H$ $\approx$ $-$24.5 mag.\ at $z$ $\sim$
0.2 (V$-$H $\approx$ 2.8). A comparison of the half-light radii of the
spheroidal components from the various data sets suggests a systematic
difference between the near-infrared and optical measurements, where
the H-band sizes are $\sim$50\% smaller than the optical sizes, but
the statistics are poor.

These shifts in colors and possibly sizes provide independent
confirmation of the presence of a young circumnuclear stellar
population in the hosts of many low-$z$ QSOs (e.g., Surace et al.\
2001; Miller \& Sheinis 2003; Canalizo et al.\ 2006, 2007; Schweitzer
et al.\ 2006; Jahnke et al.\ 2004, 2007, and references therein). A
young stellar population is a natural by-product of gas-rich galaxy
mergers. One would therefore naively expect correlations between R$-$H
and V$-$H colors and indicators of the merger phase, such as
PSF-masked residuals, PSF-to-host ratios, infrared excesses, and FIR
strength. No obvious trend is observed when combining ULIRGs and
PG~QSOs, but (1) the statistics are poor (the number of objects is
never more than 11), (2) variations in the dust content and dust
spatial distribution may be masking underlying correlations (this
possible ``cosmic conspiracy'' between stellar evolution and
extinction was also mentioned in Tacconi et al. 2002), and (3) the
host colors exclude any emission from point-source nuclear starbursts
since this nuclear emission was removed during the PSF subtraction
procedure. So the host colors of these systems need not be correlated
with the merger phase if the bulk of the emission from merger-induced
star formation is in the nuclear regions (this is the case for most if
not all ULIRGs, e.g., Soifer et al.\ 2000, and possibly also in some
PG~QSOs). These three factors may also explain the lack of any obvious
color difference with morphological class or radio-loudness.

\section{Discussion}

In Paper I, we tried to answer two important questions: (1) are
ULIRGs/QSOs elliptical galaxies in formation, and (2) are ULIRGs
related to QSOs?  Here, we revisit these questions following the same
procedure as in Paper I, but this time the QSO population is better
sampled by the new NICMOS data and near- and mid-infrared
spectroscopic data recently published by our group are used to add
important physical constraints on these objects. First, in Section
5.1, we use the FP traced by inactive spheroids to address these
issues. Next, in Section 5.2, we characterize the black hole masses
and level of black-hole driven activity likely to be taking place in
the cores of these sources.

\subsection{The Fundamental Plane}

We focus our discussion on ULIRGs and QSOs with ``pure'' elliptical
hosts, i.e., excluding the bulge + disk and ambiguous systems, to
avoid uncertainties related to the bulge/disk decomposition (e.g., Kim
et al.\ 2008a, 2008b) at the cost of reducing the sample size. Figure
8$a$ shows that ULIRGs and QSOs with elliptical hosts lie near, but
not exactly on, the photometric projection of the FP for spheroids as
traced by the K$^\prime$-band data of Pahre (1999, using
H$-$K$^\prime$ = 0.35 mag.), the $z$-band data of Bernardi et al.\
(2003; using z$-$H = 1.8 mag.) and the H-band data of Zibetti et al.\
(2002).  As found in Paper I, small ULIRG and QSO hosts are
systematically brighter than inactive spheroids of the same
size. The shift in surface brightness reaches $\sim$ 1 mag.\ for
objects with half-light radii of $\la$ 1 kpc.  This systematic trend
with half-light radii for both ULIRGs and PG~QSOs is also seen in the
linear fits through the data. The fits through the ULIRGs and PG~QSOs
(dashed and solid lines in Figure 8$a$, respectively) are
indistinguishable from each other, but they are considerably steeper
than the fit through the data of the inactive spheroids (dotted
line). Interestingly, the K-band data of Rothberg \& Joseph (2006) on
optically-selected mergers (using H-K = 0.50) show a similar shift at
small half-light radii as that of our ULIRGs and PG~QSOs. In Paper I,
we speculated that the shift to brighter magnitudes among the small
ULIRG/QSO hosts was due to excess H-band emission from a young stellar
population, but did not have the relevant data to test this statement
(see also discussion in Tacconi et al.\ 2002 and the relevant new
results of Graves et al.\ 2009 and Hopkins et al.\ 2009 and Choi et
al.\ 2009 on quiescent and UV-excess early-type galaxies,
respectively). We now revisit this issue.

In Figure 8$b$, we combine the photometric measurements of Figure 8$a$
with the stellar velocity dispersion measurements of Dasyra et al.\
(2006b, 2007) and Rothberg \& Joseph (2006) and compare the results
with the data on intermediate-size inactive spheroids from Zibetti et
al.\ (2002) and Bernardi et al.\ (2003). Here again, deviations are
seen at small half-light radii in the sense that our ULIRGs and
PG~QSOs and the optically-selected mergers of Rothberg \& Joseph
(2006) fall systematically below the FP of inactive spheroids. This
effect was noted by Rothberg \& Joseph (2006) and attributed to
differences in the effective radius and brighter surface brightness,
rather than a lower velocity dispersion; this is consistent with the
explanation of excess H-band emission from a young circumnuclear
stellar population.  Additional support for this idea comes from our
result in Section 4.5 that the colors of the PG~QSO hosts are bluer
than those of inactive spheroids of similar size.

However, if we define ``surface brightness deviation'' as the
difference between the observed surface brightness and the surface
brightness expected of a inactive spheroidal galaxy with the same
half-light radius, as determined by the linear fit through the data of
Pahre (1999), Bernardi et al.\ (2003), and Zibetti et al.\ (2002) in
Figure 8$a$, we find no obvious trend between surface brightness
deviations and R$-$H host colors (derived by combining the data of
Veilleux et al.\ 2002, 2006, Kim et al.\ 2008a, and the present
paper), contrary to what would be expected if the surface brightness
deviation was indeed due solely to excess H-band emission from a young
stellar population.  This is illustrated in Figure 9$f$. In this panel
and all others of Figure 9, ULIRGs are open symbols and PG~QSOs are
filled symbols.

The other panels of Figure 9 confirm the clear trend with half-light
radii ($a$, the probability that this correlation is fortuitous is
$P[null]$ = 0.02\%) and reveal a possible tendency for PG~QSOs with
large infrared excesses to have brighter hosts than inactive spheroids
($b$). But there is no obvious correlation between surface brightness
deviation and merger phase [as determined by the PSF-masked residuals
($c$) and the PSF-host flux ratios ($d$)] or the AGN fractional
contribution to the bolometric luminosity ($e$) derived from the {\em
  Spitzer} data of Veilleux et al.\ (2009)$\footnote{ These AGN
  contributions are calculated using six independent mid-infrared AGN
  indicators that give consistent results. The bolometric luminosities
  of ULIRGs are assumed to be 1.15 $\times$ $L$(IR), while the
  bolometric luminosities of PG~QSOs are assumed to be 7 $\times$
  $L$(5100 \AA) + $L$(IR) (Netzer et al.\ 2007). See Table 1 for a
  list of the bolometric luminosities.}$. As mentioned above, ULIRGs
and PG~QSOs show no displacement in the FP from each other.  These
results seem inconsistent at first with the idea that the surface
brightness deviations in small hosts are caused solely by excess
H-band emission from star formation. If ULIRGs are the precursors of
PG~QSOs (the {\em Spitzer} data of Veilleux et al.\ 2009 are indeed
largely consistent with this scenario), the ULIRGs should have more
star formation and therefore we naively expect that ULIRG hosts should
deviate more from the FP of inactive spheroids than PG~QSO
hosts. However, as pointed out in the last paragraph of Section 4.5.,
removal of the nuclear starbursts in these objects during the PSF
subtraction may be wiping out the expected surface brightness shift
between ULIRGs and PG~QSOs in the FP. Moreover, dust may be affecting
the observed surface brightnesses and colors, particularly in ULIRG
hosts, which are systematically redder than PG~QSO hosts (Figure 9$f$;
see also Scoville et al.\ 2000).

A closer examination of Figure 8$a$ seems to indicate that the hosts
of the more radio/X-ray luminous QSOs from Dunlop et al.\ (2003) are
systematically {\em fainter} than inactive spheroids of the same size
and fit rather well the extension to larger radii of the linear fit
through the NICMOS data on ULIRGs and PG~QSOs. If real, this result
cannot be explained by excess H-band emission from star
formation. However, a number of assumptions are made when plotting the
data points of Dunlop et al.\ on Figure 8$a$.  Following Paper I, we
used the half-light radii measured from the R-band data of Dunlop et
al.\ directly, without applying any color corrections, while the
R-band surface brightness measurements of Dunlop et al.\ were shifted
assuming R$-$H = 2.8, typical of M$_R \approx -23.5$ elliptical
systems at $z$ $\sim$ 0.2 (Lilly \& Longair 1984; Fukugita et al.\
1995; Fioc \& Rocca-Volmerange 1999; Hyv\"onen et al.\ 2007,
2008). Note that a smaller R$-$H, more in line with the average value
found for the lower luminosity PG~QSO hosts (Section 4.5 and Figure
7), would bring the data points of Dunlop et al.\ further down in
Figure 8 i.e., systematically fainter than the corresponding
spheroids. Positive R$-$H radial gradients within the hosts would
increase the H-band half-light radii, but the shift between the Dunlop
et al.\ QSOs and the inactive spheroids is too large to be explained
solely by this effect. Moreover, inactive elliptical galaxies are
usually redder near the center than on the outskirts so the color
gradients are usually negative rather than positive (e.g., Peletier et
al.\ 1990). The results from our comparisons of the QSO spheroid sizes
at V, R, and H (Section 4.5) suggest a similar negative color gradient
in QSO hosts.

Another source of uncertainty in this discussion is the exact location
of the FP at half-light radii larger than 10 kpc. The catalogs of
Bernardi et al.\ (2003) and especially Pahre (1999) contain relatively
few objects of this size so the FP is not well determined from these
data.  Also, we assumed a color correction from $z$-band to H-band for
the Bernardi et al.\ surface brightness measurements that was
independent of galaxy size (and environment); this is probably an
oversimplification (e.g., Figure 5 of Hyv\"onen et al.\ 2007 suggests
redder colors for the more luminous hosts; see also Bernardi et
al. 2006 for a discussion of a dependence on environment). Recent
compilations of FP parameters among luminous inactive and active
spheroids by Bernardi et al.\ (2006), Hamilton et al.\ (2008), and
Wolf \& Sheinis (2008) do not show any significant systematic shift
between the $r$-band properties of active and inactive spheroids. In
fact, the Hamilton et al.\ and Wolf \& Sheinis data appear to be
consistent with the FP shown in Figure 8, assuming V$-$H = 3.5 and
$r-$H = 3.0, respectively.  So one should be cautious in attaching
too much importance to the apparent shift between the Dunlop et al.\
data and the FP data of Figure 8$a$.

Nevertheless, the shift to {\em brighter} magnitudes among the small
ULIRG/QSO hosts {\em is} definitely real. The fact that this shift
does not correlate strongly with star formation/dust reddening, merger
phase, and AGN strength indicators seems to indicate that it is not
solely due to excess H-band emission from star formation.  Figure 9$d$
and 9$e$ also seem to rule out the possibility that systematic
residuals associated with the PSF fitting and removal procedure are
causing these surface brightness deviations.  At this stage, we cannot
rule out the possibility that a combination of possibly severe and
counteracting effects of population age, dust extinction and geometry,
and residual scattered emission by the central AGN/starburst is
causing this systematic shift. However, we favor a more conservative
scenario where all of these effects are relatively modest.  If the
bulk of the emission from merger-induced star formation is nuclear, as
it is known to be the case for most if not all ULIRGs and possibly
also in some PG~QSOs, then the host colors and excess H-band emission
need not be correlated with the merger phase since the bulk of this
emission was removed in the PSF subtraction procedure.

\subsection{Black Hole Masses and Accretion Rates}

The host magnitudes derived from our data can in principle be used to
derive the black hole masses in the cores of these objects, assuming
the relation between black hole mass and the mass of the spheroidal
component in normal inactive galaxy (e.g., Magorrian et al.\ 1998;
Kormendy \& Gebhardt 2001; Marconi \& Hunt 2003; H\"aring \& Rix 2004)
also applies to recent mergers. Following Paper I, we use the H-band
elliptical host magnitude -- black hole mass relation in Marconi \&
Hunt (2003), log($M_{BH}$) = $-2.80 - (0.464 \times M_H$), and deduce
photometrically derived black hole masses ranging from $\sim$ 5
$\times$ 10$^7$ (PG~0844+349) to 200 $\times$ 10$^7$ M$_\odot$
(B2~2201+31A) (Table 6). The average (median) black hole mass is
$M_{\rm BH}$ = 4.4 $\pm$ 1.0 $\times$ 10$^8$ M$_\odot$ (2.5 $\times$
10$^8$ M$_\odot$; Table 7).  This derivation neglects dust extinction
outside the nuclear regions of the hosts (which would cause an
underestimate of $M_{\rm BH}$) and the presence of recent or on-going
non-nuclear star formation  (which would have the opposite effect).

Also listed in Table 6 are the photometric black hole mass estimates
for the 1-Jy ULIRGs from Paper I and Veilleux et al. (2002), and the
black hole mass estimates for these ULIRGs and PG~QSOs derived from
three other methods, when available. The dynamical estimates are from
Dasyra et al.\ (2006a, 2006b, 2007). They are based on stellar
velocity dispersions, $\sigma_*$, measured from VLT/Keck near-infrared
spectra and the $M_{\rm BH}$ -- $\sigma_*$ relation of Tremaine et
al.\ (2002). Next, we list the black hole mass estimates for the 13
PG~QSOs from the detailed reverberation mapping study of Peterson et
al.\ (2004; updated by Bentz et al.\ 2006; 3C~273 is the only ULIRG
with a black hole mass estimate based on this method). Finally, in the
last column of these tables, we list the black hole masses of PG~QSOs
from Vestergaard \& Peterson (2006) based on the virial method. These
virial estimates are derived from the widths of the single-epoch
H$\beta$ profiles measured by Boroson \& Green (1992) and an empirical
relationship between broad-line region (BLR) size and 5100 \AA\
luminosity that is calibrated to the improved mass measurements of
nearby AGNs based on emission-line reverberation mapping.

Table 6 lists the black hole mass estimates from the four different
methods. Table 7 lists the averages, medians, and standard deviations
from the averages of the black hole mass estimates derived from each
method.  Figure 10 compares the results from the various methods on an
object-by-object basis using the data in Table 6.  Figures 10$b$ and
10$c$ indicate that the photometric, reverberation, and virial black
hole mass estimates generally agree with each other to within a factor
of $\sim$ 3 or better. On the other hand, the dynamical black hole
mass estimates in ULIRGs (PG~QSOs) are systematically smaller by a
factor of $\sim$ 7 ($\sim$ 3$-$4) on average than the other estimates.
Figures 10$a$ and 10$d$ suggest that the discrepancies between the
dynamical measurements and the photometric and reverberation mapping
measurements increase with increasing black hole masses, while Figure
10$e$ shows no obvious trend with the virial black hole masses.  Note
in passing that the large scatter in Figure 10$a$ implies that the
Faber-Jackson relation does not apply to these systems.

It is beyond the scope of this paper to try to explain the origins of
these discrepancies. Here we simply describe the principal sources of
uncertainties for each method.  The photometric method relies on the
unproven assumption that the $M_{\rm BH}$ -- $M_{\rm spheroid}$
relation of Marconi \& Hunt (2003) applies to recent mergers.  In
addition, as discussed in Section 5.1, the photometric measurements
from our data may be affected by a number of effects (non-nuclear star
formation and dust extinction, PSF subtraction) which could therefore
add uncertainties to the photometric black hole mass estimates
[similar results are found when we exclude bulge + disk systems so the
uncertainties in the bulge/disk decomposition (Kim et al.\ 2008a,
2008b) doe not appear to be a major issue here].  Note, however, that
if the surface brightness deviations seen in Figure 8$a$ are due to a
combination of these effects, then one would expect the photometric
black hole mass estimates to be overestimated in the smaller hosts
with the smaller black hole masses, the opposite of what is needed to
explain the trends of increasing discrepancies at larger black hole
masses. To further test this hypothesis we took the worst possible
scenario and assumed that the surface brightness deviations inferred
from Figure 8$a$ were due entirely to excess H-band emission from a
young stellar population and corrected the photometric black hole
masses accordingly. The results are shown by the horizontal segments
in Figures 10$a-c$. These shifts do not significantly improve the
agreement with the other methods.

The dynamical black hole mass measurements are based on two important
but largely unproven assumptions: the young star probed by the CO
observations of Dasyra et al.\ trace the full velocity dispersion of
the spheroid and the $M_{\rm BH}$ -- $\sigma_*$ relation(s) apply to
recent mergers. Recent simulations (e.g., Dasyra et al.\ 2006b;
Johansson et al.\ 2009) provide support for this last assumption, but
it is far from being the final word given the difficulty in modeling
the complex processes associated with star formation and black hole
growth on sub-pc to kpc scales. The first assumption has been
discussed in the context of optically-selected merger remnants, where
Ca II triplet velocity dispersion measurements are found to be
systematically larger than CO measurements by a factor of up to $\sim$
2 (Rothberg \& Joseph 2006; Dasyra et al.\ 2006b; Rothberg 2009; see
also Silge \& Gebhardt 2003 in elliptical galaxies). The
dynamically-derived black hole masses scale with the fourth power of
the velocity dispersions, so this systematic shift between optical and
near-infrared measurements could conceivably explain some of the
discrepancy between the dynamical measurements and the other
measurements. 

Finally, the reverberation mapping and virial measurements are widely
considered to be the most reliable estimates of black hole masses.
However, they too are subject to possibly significant
uncertainties. In particular, the scale factor $f$, which accounts for
the unknown geometry, kinematics, inclination of the broad-line
region, may depend on luminositiy and accretion rate (e.g., Collin et
al.\ 2006). The value adopted by Peterson et al.\ (2004) and
Vestergaard \& Peterson (2006), $f = 5.5$, was derived from lower
luminosity AGN (Onken et al.\ 2004) and may not apply to the higher
luminosity PG~QSOs of our sample (see Dasyra et al.\ 2007 and Watson
et al.\ 2008 for a more detailed discussion of the origins of the
discrepancies between the dynamical and reverberation mapping
methods).

Given the substantial uncertainties affecting all of the black hole
mass measurements, it is in fact remarkable that a large subset of
these measurements agree with each other to within of $\sim$ 3 or
better.  In the following discussion, we adopt our photometric black
hole mass estimates at face value, keeping in mind of the possibly
large uncertainties on these black hole mass estimates, and derive the
Eddington ratio i.e.,  the ratio of AGN bolometric luminosity to
the Eddington luminosity, $L_{\rm Edd} = 3.3 \times 10^4 M_{\rm
  BH}/M_\odot$ $L_\odot$, for each system. This ratio is an objective
indicator of the level of nuclear activity in these systems. The AGN
fractional contributions to the bolometric luminosities of the PG~QSOs
and ULIRGs are taken directly from our {\em Spitzer} study (Veilleux
et al.\ 2009; see details in footnote \#5 in Section 5.1).  Some of the
results have already been discussed in Veilleux et al.\ (2009) and are
not repeated here. Figure 11 focuses exclusively on the ULIRGs and
radio-quiet and radio-loud PG~QSOs in the {\em HST} sample. These
three classes of objects have statistically the same photometric
Eddington ratios, of order $\sim$ 3-30\% ($\sim$10\% on average). This
result is similar to those derived by McLeod \& McLeod
(2001). Interestingly, none of the ULIRGs and PG~QSOs in our sample
require super-Eddington mass accretion rates.  This remains true for
all but two objects after we correct the spheroid host magnitudes for
possible excess H-band emission from young stellar population
(indicated by the horizontal segments in Figure 11). The corrected
Eddington ratios are then $\sim$ 30\% on average.

\section{Conclusions}

As part of QUEST, we have supplemented our original {\em HST} NICMOS
H-band imaging data set on 7 PG~QSOs from Paper I with an additional
set of 21 PG~QSOs, for a total of 28 objects. The results from our
detailed two-dimensional analysis of this larger PG~QSO sample were
then compared with the data from Paper I on ULIRGs, which were
analyzed exactly in the same way, and those from literature. The main
conclusions of our study are the followings:

\begin{itemize}

\item The majority (57\%) of the PG~QSOs show signs of a recent
  galactic interaction.

\item Eleven (39\%) PG~QSOs show a prominent elliptical morphology,
  nine (32\%) have a distinct stellar disk in addition to a central
  bulge, and the others have a morphology that is ambiguous either due
  to severe merger-induced disturbances (5/28, 18\%) or mismatch in
  the point-spread function (3/28, 11\%).  .

\item The fraction of QSOs with elliptical host is larger among QSOs
  with undetected FIR emission, small infrared excess, and luminous
  hosts.

\item The hosts of FIR-bright QSOs show a tendency to have more
  pronounced merger-induced morphological anomalies and smaller
  QSO-to-host luminosity ratios on average than the hosts of FIR-faint
  QSOs.

\item The host sizes and luminosities of the radio-quiet (radio-loud)
  PG~QSOs in our sample are statistically indistinguishable from
  (larger than) those of the 26 highly-nucleated ULIRG hosts presented
  in Paper I.  ULIRGs, radio-quiet PG~QSOs, and radio-loud PG~QSOs
  with elliptical hosts lie close to, but not exactly on, the FP of
  inactive spheroids. We confirm the tendency noted in Paper I for
  objects with small ($\la$ 2 kpc) spheroids to be up to 1 mag.\
  brighter than normal inactive spheroids.  Comparisons of our H-band
  host magnitudes and sizes with similar R- and V-band data taken from
  the literature support the existence of a young stellar population
  outside the nuclear region of several PG~QSOs and ULIRGs which may
  contribute to the observed excess H-band emission. However, no
  obvious trend is seen between this excess H-band emission and host
  R$-$H color, merger phase, or AGN indicators in ULIRGs and PG~QSOs,
  suggesting that other effects like dust extinction are also at
  play. PSF subtraction may also wipe out correlations with merger
  phase in systems with strong merger-induced nuclear starbursts
  (i.e., most ULIRGs and possibly some PG~QSOs).

\item The H-band spheroid-host luminosities of the PG~QSOs,
  uncorrected for extinction or the presence of a young stellar
  population, imply black hole masses ranging from $\sim$ 5 to 200
  $\times$ 10$^7$ M$_\odot$. These values are similar to those of the
  ULIRGs from Paper I, within a factor of $\sim$3 from black hole mass
  estimates based on the reverberation mapping and virial methods, but
  significantly larger than those derived from the stellar velocity
  dispersion method. These discrepancies are arguably within the range
  of the large uncertainties on all these measurements. 

\item Sub-Eddington mass accretion rates of order $\sim$ 3$-$30\% are
  implied for all PG~QSOs and ULIRGs in our sample when the
  photometric black hole mass estimates are combined with our
  published {\em Spitzer} estimates of the AGN contributions to the
  bolometric luminosities in these objects. Corrections due to
  possible excess H-band emission from a young circumnuclear stellar
  population increase the average mass accretion rate by a factor of
  $\sim$ 3.

\end{itemize}

By and large, these results and those of Paper I support the merger
scenario where QSO activity of moderate luminosity is triggered by
major galaxy mergers that result in the formation of intermediate-mass
spheroids. The weaker merger-induced morphological anomalies found
among Seyfert-like ULIRGs (Paper I) and PG~QSOs with elliptical hosts
and small infrared excess indicate that nuclear activity is indeed
seen preferentially in late-stage mergers. The disk components,
detected in all QSOs with AGN bolometric luminosities less than $\sim$
10$^{11.5}$ L$_\odot$, can be explained in this merger scenario if
substantial and rapid accretion of residual gas took place after the
merger.

\vskip 0.1in

\acknowledgements S.V., D.C.K., and D.S.N.R.\ were supported in part
by NASA through grant HST-GO-10906.01-A. S. V. acknowledges support
from a Senior Award from the Alexander von Humboldt Foundation and
thanks the host institution, MPE Garching, where some of this work was
performed.  C.Y.P.\ is grateful to Space Telescope Science Institute
for support through the Institute Fellowship program and to the
National Research Council of Canada through the Plaskett Fellowship
program at the Herzberg Institute of Astrophysics. We thank R.\
Davies, K.\ Jahnke, and E.\ Bell for comments on an earlier version of
the manuscript, and the anonymous referee for a thoughtful
review. This work has made use of NASA's Astrophysics Data System
Abstract Service and the NASA/IPAC Extragalactic Database (NED), which
is operated by the Jet Propulsion Laboratory, California Institute of
Technology, under contract with the National Aeronautics and Space
Administration.

\clearpage

\begin{deluxetable}{lcccrcrrrccl}
\tabletypesize{\scriptsize}
\tablenum{1}
\tablewidth{500pt}
\tablecaption{Sample}
\tablehead{
\multicolumn{1}{c}{Name} &
\multicolumn{1}{c}{Other Name} &
\multicolumn{1}{c}{$z$} &
\multicolumn{1}{c}{log$\left(L_{\rm B} \over L_\odot \right)$} &
\multicolumn{1}{c}{log$\left(L_{\rm IR} \over L_\odot \right)$} &
\multicolumn{1}{c}{log$\left(L_{\rm BOL} \over L_\odot \right)$} &
\multicolumn{1}{c}{$L_{\rm IR} \over L_{\rm B}$} &
\multicolumn{1}{c}{$L_{\rm IR} \over L_{\rm BOL}$} &
\multicolumn{1}{c}{$f_{25} \over f_{60}$} &
\multicolumn{1}{c}{Radio} &
\multicolumn{1}{c}{FIR Class} \\
\multicolumn{1}{c}{(1)} &
\multicolumn{1}{c}{(2)} &
\multicolumn{1}{c}{(3)} &
\multicolumn{1}{c}{(4)} &
\multicolumn{1}{c}{(5)} &
\multicolumn{1}{c}{(6)} &
\multicolumn{1}{c}{(7)} &
\multicolumn{1}{c}{(8)} &
\multicolumn{1}{c}{(9)} &
\multicolumn{1}{c}{(10)} &
\multicolumn{1}{c}{(11)}
}
\startdata
PG0007$+$106   & III Zw 2   & 0.089 & -22.28 &   11.44 &  12.23 &   24.6  &   0.16 & 0.765  & L & Weak       &\\    
PG0026$+$129   & \nodata    & 0.142 & -23.84 &$<$11.47 &  12.07 & $<$3.31 &$<$0.25 &$<$0.410& Q & Undetected &\\    
PG0050$+$124   & I Zw 1     & 0.061 & -22.68 &   11.87 &  12.07 &   29.5  &   0.63 & 0.382  & Q & Strong     &\\    
PG0157$+$001   & Mrk 1014   & 0.163 & -24.57 &   12.63 &  12.69 &   44.7  &   0.87 & 0.243  & Q & Strong     &\\    
PG0838$+$770   & VII Zw 244 & 0.131 & -21.55 &   11.45 &  11.76 &   12.9  &   0.49 & 0.455  & Q & Weak       &\\    
PG0844$+$349   & \nodata    & 0.064 & -22.19 &   10.93 &  11.44 &    2.14 &   0.31 & 0.628  & Q & Weak       &\\    
PG0923$+$201   & \nodata    & 0.190 & -24.23 &$<$12.12 &  12.45 &$<$20.9  &$<$0.47 &$<$0.322& Q & Undetected &\\    
PG1001$+$054   & \nodata    & 0.161 & -22.94 &   11.56 &  11.86 &    9.12 &   0.50 & 0.456  & Q & Strong     &\\
\enddata
\tablenotetext{\ }
{{\em Col 1:}\ Object name.}
\tablenotetext{\ }
{{\em Col 2:}\ Other name.}
\tablenotetext{\ }
{{\em Col 3:}\ Redshift.}
\tablenotetext{\ }
{{\em Col 4:}\ Blue luminosity.}
\tablenotetext{\ }
{{\em Col 5:}\ Infrared luminosity calculated from prescription of Sanders \& Mirabel (1996).}
\tablenotetext{\ }
{{\em Col 6:}\ Bolometric luminosity calculated from 7 $\times$ L(5100$\AA$) + L(IR) (Netzer et al.\ 2007).}
\tablenotetext{\ }
{{\em Col 7:}\ Ratio of infrared to blue luminosities.}
\tablenotetext{\ }
{{\em Col 8:}\ Ratio of infrared to bolometric luminosities.}
\tablenotetext{\ }
{{\em Col 9:}\ $f_{25}/f_{60}$ flux ratio. Object with colon is uncertain.}
\tablenotetext{\ }
{{\em Col 10:}\ Radio loudness (Q: radio quiet, L: radio loud) from Kellermann et al (1994) except for two objects (B2 2201$+$31A and PG2349$-$014 from Becker et al., 1991).}
\tablenotetext{\ }
{{\em Col 11:}\ FIR strength according to Netzer et al. (2007).}
\end{deluxetable}

\clearpage

\begin{deluxetable}{lrrrrrrrrrrrrr}
\tabletypesize{\scriptsize}
\tablenum{2}
\tablewidth{450pt}
\tablecaption{One Galaxy Component Fits}
\tablehead{
\multicolumn{1}{c}{Name} &
\multicolumn{1}{c}{n} &
\multicolumn{1}{c}{r$_{1/2}$} &
\multicolumn{1}{c}{b/a} &
\multicolumn{1}{c}{PA} &
\multicolumn{1}{c}{m$_n$} &
\multicolumn{1}{c}{M$_n$} &
\multicolumn{1}{c}{m$_{PSF}$} &
\multicolumn{1}{c}{M$_{PSF}$} &
\multicolumn{1}{c}{R$_1$} &
\multicolumn{1}{c}{R$_2$} &
\multicolumn{1}{c}{$\chi^2_{\nu1}$} &
\multicolumn{1}{c}{$\chi^2_{\nu2}$} \\
\multicolumn{1}{c}{(1)} &
\multicolumn{1}{c}{(2)} &
\multicolumn{1}{c}{(3)} &
\multicolumn{1}{c}{(4)} &
\multicolumn{1}{c}{(5)} &
\multicolumn{1}{c}{(6)} &
\multicolumn{1}{c}{(7)} &
\multicolumn{1}{c}{(8)} &
\multicolumn{1}{c}{(9)} &
\multicolumn{1}{c}{(10)} &
\multicolumn{1}{c}{(11)} &
\multicolumn{1}{c}{(12)} &
\multicolumn{1}{c}{(13)}
}
\startdata
PG0007$+$106   &   1.0 &   1.79 &   0.93 &  134.4  &  14.47 &$-$23.59 &  13.58 &$-$24.48 &  64.2 &  45.7 &   8.6 &   1.5 & \\    
               &   4.0 &   2.97 &   0.92 &  137.9  &  13.88 &$-$24.18 &  13.63 &$-$24.43 &  36.9 &  22.9 &   6.2 &   0.5 & \\    
               &   7.2 &   5.33 &   0.92 &  137.4  &  13.53 &$-$24.53 &  13.67 &$-$24.39 &  35.2 &  22.0 &   6.0 &   0.5 & \\    
PG0026$+$129   &   1.0 &   2.07 &   0.85 &$-$64.9  &  15.74 &$-$23.40 &  13.40 &$-$25.74 &  92.9 &   9.3 &  27.6 &   1.3 & \\    
               &   4.0 &   0.90 &   0.83 &  111.6  &  15.14 &$-$24.00 &  13.47 &$-$25.67 &  69.8 &   7.0 &  26.4 &   1.3 & \\    
               &   5.5 &   0.95 &   0.80 &  110.3  &  15.16 &$-$23.98 &  13.46 &$-$25.68 &  71.1 &   6.9 &  26.3 &   1.3 & \\    
PG0050$+$124$^{(a)}$ &   1.0 &   2.38 &   0.82 &   30.6  &  12.77 &$-$24.42 &  12.12 &$-$25.07 &  55.7 &  15.1 & 234.3 &  86.7 & \\    
               &   4.0 &   1.64 &   0.83 &   35.2  &  12.36 &$-$24.83 &  12.25 &$-$24.94 &  37.5 &  10.5 & 163.9 &  68.6 & \\
\enddata
\tablenotetext{\ }
{{\it Col 1:}\ Object name.}
\tablenotetext{\ }
{{\it Col 2:}\ S\'ersic index.}
\tablenotetext{\ }
{{\it Col 3:}\ Half-light radius in kpc of S\'ersic component.}
\tablenotetext{\ }
{{\it Col 4:}\ Axis ratio of S\'ersic component.}
\tablenotetext{\ }
{{\it Col 5:}\ Position angle (East of North) of major axis of S\'ersic component.}
\tablenotetext{\ }
{{\it Col 6:}\ Apparent H magnitude of S\'ersic component.}
\tablenotetext{\ }
{{\it Col 7:}\ Absolute H magnitude of S\'ersic component.}
\tablenotetext{\ }
{{\it Col 8:}\ Apparent H magnitude of PSF component.}
\tablenotetext{\ }
{{\it Col 9:}\ Absolute H magnitude of PSF component.}
\tablenotetext{\ }
{{\it Col 10:}\ Absolute residuals normalized to total host galaxy flux (\%). See Section 4.3 for more detail.}
\tablenotetext{\ }
{{\it Col 11:}\ PSF-masked absolute residuals normalized to total host galaxy
flux (\%). The central PSF region brighter than 11 H mag arcsec$^{-2}$ was
masked for these calculations. See Section 4.3 for more detail.}
\tablenotetext{\ }
{{\it Col 12:}\ Reduced $\chi^2$ value.}
\tablenotetext{\ }
{{\it Col 13:}\ PSF-masked reduced $\chi^2$ value. The central PSF region  
brighter than 11 H mag arcsec$^{-2}$ was masked for these calculations.}
\tablenotetext{\ }
{$^{(a)}$ Entries for this object are uncertain because the host galaxy 
fills most of the field of view so the sky background is not well determined.}
\end{deluxetable}

\clearpage

\centering
\begin{deluxetable}{lrrrrrrrrrrrrr}
\tabletypesize{\scriptsize}
\tablenum{3}
\tablewidth{450pt}
\tablecaption{Two Galaxy Components Fits}
\bigskip
\tablehead{
\multicolumn{1}{c}{Name} &
\multicolumn{1}{c}{n} &
\multicolumn{1}{c}{r$_{\frac{1}{2}}$} &
\multicolumn{1}{c}{b/a} &
\multicolumn{1}{c}{PA} &
\multicolumn{1}{c}{m$_n$} &
\multicolumn{1}{c}{M$_n$} &
\multicolumn{1}{c}{m$_{PSF}$} &
\multicolumn{1}{c}{M$_{PSF}$} &
\multicolumn{1}{c}{R$_1$} &
\multicolumn{1}{c}{R$_2$} &
\multicolumn{1}{c}{$\chi^2_{\nu1}$} &
\multicolumn{1}{c}{$\chi^2_{\nu2}$} \\
\multicolumn{1}{c}{(1)} &
\multicolumn{1}{c}{(2)} &
\multicolumn{1}{c}{(3)} &
\multicolumn{1}{c}{(4)} &
\multicolumn{1}{c}{(5)} &
\multicolumn{1}{c}{(6)} &
\multicolumn{1}{c}{(7)} &
\multicolumn{1}{c}{(8)} &
\multicolumn{1}{c}{(9)} &
\multicolumn{1}{c}{(10)} &
\multicolumn{1}{c}{(11)} &
\multicolumn{1}{c}{(12)} &
\multicolumn{1}{c}{(13)} 
}
\startdata
PG0007$+$106   &   1.0 &   5.28 &   0.84 &    148.2 &  14.67 &$-$23.39 &  13.67 &$-$24.39 &  33.6 &  20.5 &   6.0 &   0.4 & \\ 
               &   4.0 &   0.93 &   0.93 &    127.2 &  14.61 &$-$23.45 &        &         &       &       &       &       & \\ 
PG0026$+$129   &   1.0 &   4.92 &   0.85 &     77.6 &  15.98 &$-$23.16 &  13.47 &$-$25.67 &  65.5 &   4.1 &  25.7 &   1.2 & \\ 
               &   4.0 &   0.35 &   0.80 &  $-$65.0 &  15.46 &$-$23.68 &        &         &       &       &       &       & \\ 
PG0050$+$124$^{(a)}$ &   1.0 &   3.35 &   0.88 &     37.5 &  13.76 &$-$23.43 &  12.26 &$-$24.93 &  36.4 &   9.9 & 154.5 &  59.8 & \\ 
               &   4.0 &   1.03 &   0.79 &     34.5 &  12.76 &$-$24.43 &        &         &       &       &       &       & \\ 
PG0157$+$001   &   1.0 &   9.92 &   0.89 &   $-$6.6 &  14.59 &$-$24.88 &  13.95 &$-$25.52 &  31.5 &  20.8 &   6.0 &   1.1 & \\ 
               &   4.0 &   0.90 &   0.90 &  $-$37.4 &  14.05 &$-$25.42 &        &         &       &       &       &       & \\
\enddata
\tablenotetext{\ }
{{\it Col 1:}\ Object name.}
\tablenotetext{\ }
{{\it Col 2:}\ S\'ersic index.}
\tablenotetext{\ }
{{\it Col 3:}\ Half-light radius in kpc of S\'ersic component.}
\tablenotetext{\ }
{{\it Col 4:}\ Axis ratio of S\'ersic component.}
\tablenotetext{\ }
{{\it Col 5:}\ Position angle (East of North) of major axis of S\'ersic component.}
\tablenotetext{\ }
{{\it Col 6:}\ Apparent H magnitude of S\'ersic component.}
\tablenotetext{\ }
{{\it Col 7:}\ Absolute H magnitude of S\'ersic component.}
\tablenotetext{\ }
{{\it Col 8:}\ Apparent H magnitude of PSF component.}
\tablenotetext{\ }
{{\it Col 9:}\ Absolute H magnitude of PSF component.}
\tablenotetext{\ }
{{\it Col 10:}\ Absolute residuals normalized to total host galaxy flux (\%). See Section 4.3 for more detail.}
\tablenotetext{\ }
{{\it Col 11:}\ PSF-masked absolute residuals normalized to total host galaxy
flux (\%). The central PSF region brighter than 11 H mag arcsec$^{-2}$ was
masked for these calculations. See Section 4.3 for more detail.}
\tablenotetext{\ }
{{\it Col 12:}\ Reduced $\chi^2$ value.}
\tablenotetext{\ }
{{\it Col 13:}\ PSF-masked reduced $\chi^2$ value. The central PSF region  
brighter than 11 H mag arcsec$^{-2}$ was masked for these calculations.}
\tablenotetext{\ }
{$^{(a)}$ Entries for this object are uncertain because the host galaxy
fills most of the field of view so the sky background is not well determined.}
\end{deluxetable}

\clearpage

\centering
\begin{deluxetable}{lccccc}
\small
\tablenum{4}
\tablewidth{300pt}
\tablecaption{Details of PG QSO Small Structure}
\bigskip
\tablehead{
\multicolumn{1}{c}{Name} &
\multicolumn{1}{c}{$\Delta$ R.A.} &
\multicolumn{1}{c}{$\Delta$ Dec.} &
\multicolumn{1}{c}{r$_{\frac{1}{2}}$} &
\multicolumn{1}{c}{m$_H$} \\
\multicolumn{1}{c}{(1)} &
\multicolumn{1}{c}{(2)} &
\multicolumn{1}{c}{(3)} &
\multicolumn{1}{c}{(4)} &
\multicolumn{1}{c}{(5)}  
}
\startdata
PG1119$+$120   &   -0.6 &    2.8 &   0.87 &  16.00& \\ 
PG1302$-$102 \#1&   -0.6 &    0.9 &   0.51 &  16.96& \\ 
PG1302$-$102 \#2&    0.0 &    2.4 &   0.84 &  17.91& \\ 
PG1411$+$442   &   -1.2 &   -1.8 &   0.57 &  16.81& \\ 
PG1426$+$015   &    1.3 &   -2.1 &   0.41 &  15.93& \\ 
PG1613$+$658   &   -2.2 &    0.8 &   0.76 &  15.52& \\ 
PG2349$-$014   &    1.9 &   -0.4 &   1.13 &  18.55& \\ 
\enddata
\tablenotetext{\ }
{{\it Col 1:}\ Object name.}
\tablenotetext{\ }
{{\it Col 2 - Col 3:}\ Offsets relative to the QSO nucleus.}
\tablenotetext{\ }
{{\it Col 4:}\ Half-light radius in kpc of S\'ersic component.}
\tablenotetext{\ }
{{\it Col 5:}\ Apparent H magnitude of component.}
\end{deluxetable}

\clearpage

\begin{deluxetable}{llrrccccclllccc}
\tabletypesize{\tiny}
\tablenum{5}
\tablecaption{Summary$^{(a)}$}
\tablehead{
\multicolumn{9}{c}{} &
\multicolumn{3}{c}{Visual Features} &
\multicolumn{2}{c}{} \\
\cline{10-12}
\multicolumn{1}{c}{Name} &
\multicolumn{1}{c}{M$_{total}$} &
\multicolumn{1}{c}{M$_{PSF}$} &
\multicolumn{1}{c}{M$_{host}$} &
\multicolumn{1}{c}{M$_{model}$} &
\multicolumn{1}{c}{${\rm I}_{PSF} \over {\rm I}_{host}$} &
\multicolumn{1}{c}{${\rm I}_{model} \over {\rm I}_{host}$} &
\multicolumn{1}{c}{r$_{\frac{1}{2}}$} &
\multicolumn{1}{c}{$<\mu_{\frac{1}{2}}>$} &
\multicolumn{1}{c}{Bar} &
\multicolumn{1}{c}{Arms} &
\multicolumn{1}{c}{Dis.} &
\multicolumn{1}{c}{MC} &
\multicolumn{1}{c}{IC} \\
\multicolumn{1}{c}{(1)} &
\multicolumn{1}{c}{(2)} &
\multicolumn{1}{c}{(3)} &
\multicolumn{1}{c}{(4)} &
\multicolumn{1}{c}{(5)} &
\multicolumn{1}{c}{(6)} &
\multicolumn{1}{c}{(7)} &
\multicolumn{1}{c}{(8)} &
\multicolumn{1}{c}{(9)} &
\multicolumn{1}{c}{(10)} &
\multicolumn{1}{c}{(11)} &
\multicolumn{1}{c}{(12)} &
\multicolumn{1}{c}{(13)} &
\multicolumn{1}{c}{(14)}
}
\startdata
PG0007$+$106   &  -25.03 &  -24.43 &  -24.11 & -24.18 &   1.34 &   1.07 &   2.97 &  17.13 &   No &  Yes &  Yes  &  E    & IVb & \\
PG0026$+$129   &  -25.92 &  -25.67 &  -24.21 & -24.00 &   3.84 &   0.82 &   0.90 &  14.92 &   No &   No &  Yes  &  E    &  V  & \\
PG0050$+$124$^{(b)}$ &  -25.62 &  -24.93 &  -24.81 & -24.79 &   1.12 &   0.99 &   1.03 &  14.45 &   No &  Yes &  Str  &  B+D  & IVb & \\
PG0157$+$001   &  -26.48 &  -25.61 &  -25.84 & -25.65 &   0.80 &   0.83 &   1.82 &  14.88 &   No &  Yes &  Str  &  E    & IVb & \\
PG0838$+$770$^{(b)}$ &  -25.21 &  -23.80 &  -24.87 & -24.90 &   0.37 &   1.03 &   0.56 &  14.45 &  Yes &  Yes &  Yes  &  B+D  & IVb & \\
PG0844$+$349   &  -24.73 &  -24.17 &  -23.74 & -23.72 &   1.50 &   0.98 &   0.28 &  13.40 &   No &  Yes &  Yes  &  B+D  & IVb & \\
PG0923$+$201   &  -26.34 &  -26.05 &  -24.74 & -24.88 &   3.34 &   1.14 &   1.30 &  15.01 &   No &   No &   No  &  E    &  V  & \\
PG1001$+$054   &  -25.34 &  -25.12 &  -23.49 & -23.67 &   4.50 &   1.18 &   2.38 &  17.43 &  Yes?& Yes? & Yes?  &  E    &  V  & \\
\enddata
\tablenotetext{\ }
{{\it Col 1:}\ Object name.}
\tablenotetext{\ }
{{\it Col 2:}\ Total absolute magnitude (host + PSF - companions).}
\tablenotetext{\ }
{{\it Col 3:}\ Absolute magnitude of PSF component.}
\tablenotetext{\ }
{{\it Col 4:}\ Absolute magnitude of host galaxy (including tidal features).}
\tablenotetext{\ }
{{\it Col 5:}\ Absolute magnitude of best-fitting galaxy host model.}
\tablenotetext{\ }
{{\it Col 6:}\ PSF-to-host intensity ratio.}
\tablenotetext{\ }
{{\it Col 7:}\ Model-to-host intensity ratio.}
\tablenotetext{\ }
{{\it Col 8:}\ Half-light radius in kpc of n=4 S\'ersic component.}
\tablenotetext{\ }
{{\it Col 9:}\ Mean surface brightness of n=4 S\'ersic component within half-light radius in 
H mag.\ arcsec$^{-2}$.}
\tablenotetext{\ }
{{\it Col 10:}\ Presence of a stellar bar.}
\tablenotetext{\ }
{{\it Col 11:}\ Presence of spiral arms.}
\tablenotetext{\ }
{{\it Col 12:}\ Presence of merger-induced disturbance (Str=strongly disturbed). }
\tablenotetext{\ }
{{\it Col 13:}\ Morphological class: E = elliptical, B+D = bulge+disk, D = disk, A = ambiguous.
Question marks (``?'') indicate uncertain classification due to PSF mismatch (see Sections 3 and 4.1). }
\tablenotetext{\ }
{{\it Col 14:}\ Interaction class
(see VSK02 for details on the definitions).}  
\tablenotetext{\ }
{}
\tablenotetext{\ }
{$^{(a)}$ Entries in this table are the parameters from the
  best-fitting one or two galaxy component models (Tables 2 and 3).
}
\tablenotetext{\ }
{$^{(b)}$ Entries for this object are uncertain because the host galaxy
fills most of the field of view so the sky background is not well determined.}
\end{deluxetable}

\begin{deluxetable}{lrrrrc}
\tablenum{6}
\tablecolumns{6}
\tabletypesize{\scriptsize}
\tablecaption{Black Hole Mass}
\tablewidth{0pt}
\tablehead{
\colhead{} & \multicolumn{4}{c}{$M_{BH}/10^7 M_\sun$} & \colhead{} \\
\colhead{Galaxy} & \colhead{phot} & \colhead{dyn} & \colhead{rev} & \colhead{vir} & \colhead{References} \\
\colhead{(1)} & \colhead{(2)} & \colhead{(3)} & \colhead{(4)} & \colhead{(5)} & \colhead{(6)}
}
\startdata
\cutinhead{ULIRGs}
IRAS00091$-$0738    & 22.1 &   2.46  &\nodata &\nodata&     2 1 . . \\
IRAS00188$-$0856    & 26.6 & \nodata &\nodata &\nodata&     3 . . . \\
IRAS00199$-$7426    &  \nodata&   2.95  &\nodata &\nodata&     . 4 . . \\
IRAS00262$+$4251    &  \nodata&   7.02  &\nodata &\nodata&     . 1 . . \\
IRAS00397$-$1312    & 31.5 &   1.05  &\nodata &\nodata&     3 1 . . \\
IRAS00456$-$2904SW  & 27.7 &   5.79  &\nodata &\nodata&     3 1 . . \\
IRAS00456$-$2904NE  &  2.1 & \nodata &\nodata &\nodata&     3 . . . \\
IRAS00482$-$2721    & 10.6 & \nodata &\nodata &\nodata&     2 . . . \\
\enddata
\tablerefs{1 = Dasyra et al. 2006b; 2 = Veilleux et al. 2002; 3 = Veilleux et al. 2006; 4 = Dasyra et al. 2006a; 5 = This paper; 6 = Peterson et al. 2004; 7 = Dasyra et al. 2007; 8 = Vestergaard \& Peterson, 2006}
\tablecomments{
Col.(1): Galaxy name.  Col.(2-5): Black hole mass, in units of 10$^7$ $M_\sun$, as determined from spheroid luminosity, spheroid velocity dispersion, reverberation mapping, and virial relation.  Col.(6): $M_{BH}$ references.
}
\end{deluxetable}

\begin{deluxetable}{crrrr}
\tablenum{7}
\tablecolumns{5}
\tabletypesize{\scriptsize}
\tablecaption{Black Hole Mass Statistics}
\tablewidth{0pt}
\tablehead{
\colhead{} & \multicolumn{4}{c}{log($M_{BH}/10^7 M_\sun$)} \\
\colhead{Quantity} & \colhead{phot} & \colhead{dyn} & \colhead{rev} & \colhead{vir} \\
\colhead{(1)} & \colhead{(2)} & \colhead{(3)} & \colhead{(4)} & \colhead{(5)} \\
}
\startdata
\cutinhead{ULIRGs}
  No. Gals.  &  134   & 72    &   1   &    0 \\
  Average    & 42.85  & 5.76  & 88.60 &\nodata \\
  Median     & 26.90  & 3.76  & 88.60 &\nodata \\
  Std. Dev.  & 96.84  & 7.49  & 88.60 &\nodata \\
\cutinhead{PG QSOs}
  No. Gals.  & 27     & 10    & 14   &   18  \\
  Average    & 43.77  & 10.61 & 42.71& 34.07 \\
  Median     & 24.90  & 9.66  & 41.70& 18.20 \\
  Std. Dev.  & 48.59  & 6.34  & 34.15& 46.52 \\
\cutinhead{ULIRGs $+$ PG QSOs}
  No. Gals.  & 161    & 82    & 15   &   18  \\
  Average    & 43.01  & 6.35  & 45.77& 34.07 \\
  Median     & 25.20  & 4.19  & 44.00& 18.20 \\
  Std. Dev.  & 90.43  & 7.50  & 34.98& 46.52 \\
\enddata
\tablecomments{
Col.(1): Quantity.  Col.($2-5$): Black hole mass statistics, where individual $M_{BH}$ values are first divided by 10$^7$ $M_\sun$ and then logged.  Masses are determined from spheroid luminosity, spheroid velocity dispersion, reverberation mapping, and virial relation.
}
\end{deluxetable}

\clearpage

\begin{figure*}[ht]
\epsscale{0.8}
\caption{ Results from the GALFIT one galaxy component analysis. For
  each object, panel ($a$) shows the original data while the other
  panels show the residuals after subtracting three different models:
  ($b$) PSF + S\'ersic component with $n$ = 1 (exponential disk),
  ($c$) PSF + S\'ersic component with $n$ = 4 (de Vaucouleurs
  spheroid), and ($d$) PSF + S\'ersic component with unconstrained
  index. The intensity scale is logarithmic and the horizontal segment
  between panels ($b$) and ($c$) represents 10  kpc. The tickmarks in
  the panels are separated by 5\arcsec. }
\end{figure*}

%
%
%
%
%

\begin{figure*}[ht]
\epsscale{0.7}
\caption{ Results from the GALFIT two galaxy component analysis for 16
   PG~QSOs with possible low-surface-brightness exponential disks. A
   disk is detected unambiguously in 9 of these objects: PG~0050+124,
   0838+770, 0844+349, 1119+120, 1126-041, 1229+204, 1426+015,
   1440+356, and 2130+099 (see details in Table 5). In the other
   systems presented here, the addition of a disk component did not
   improve the fits significantly.  Panel ($a$) shows the original
   data and panel ($b$) shows the residuals after subtracting a model
   with a PSF, a bulge-like S\'ersic component with $n$ = 4, and a
   disk-like S\'ersic component with $n$ = 1. Panels ($c$) and ($d$)
   show the surface brightness distributions of the two S\'ersic
   components used in the model. The centroids of the components are
   left unconstrained. The intensity scale is logarithmic and the
   vertical segment between panels ($b$) and ($c$) represents 10
   kpc. The tickmarks in each panel are separated by 5\arcsec.}
\end{figure*}

%
%

\clearpage

\begin{figure*}[ht]
\epsscale{1.0}
\plotone{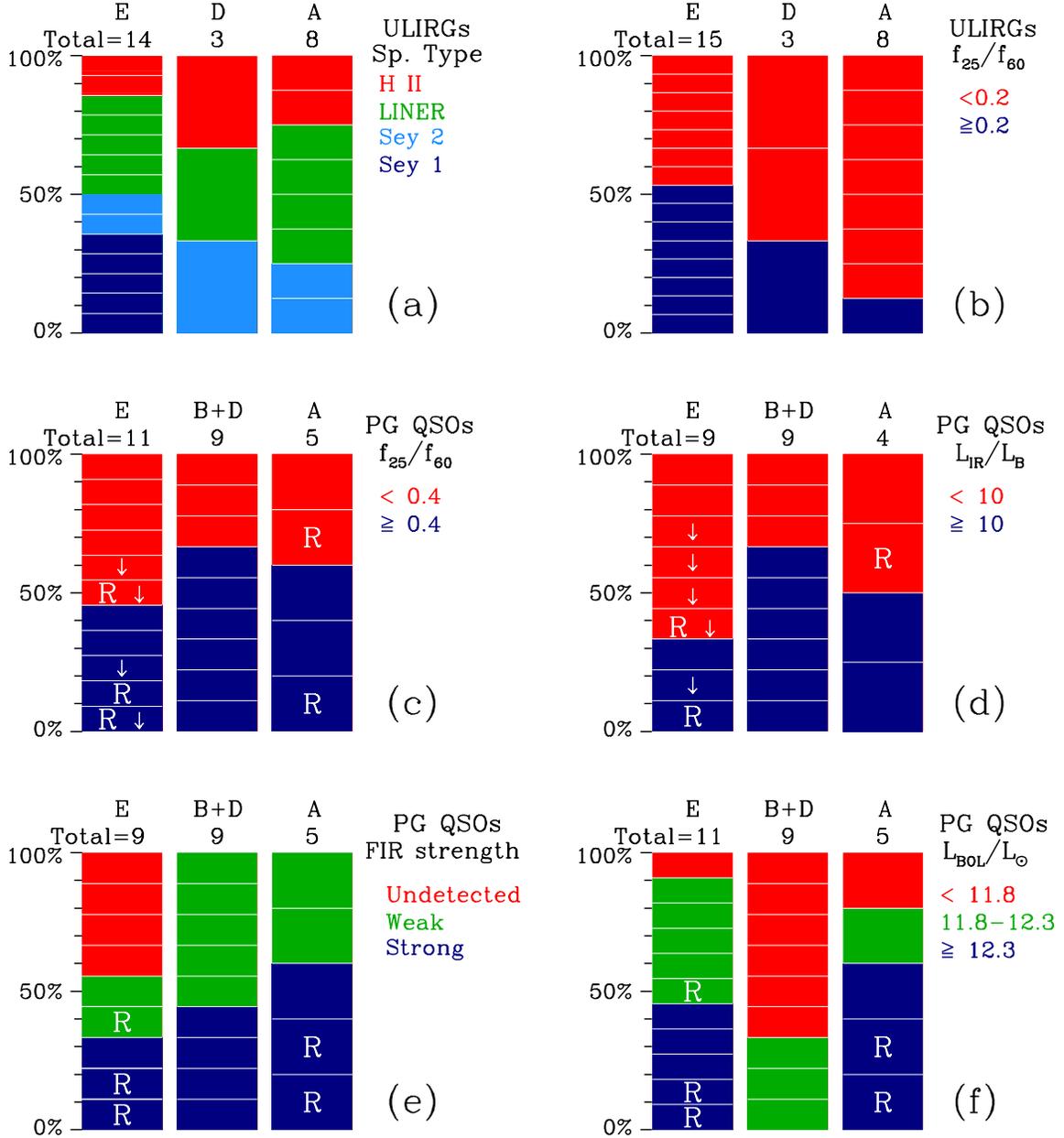}
\caption{ Trends between morphological types (E = elliptical, D =
  disk, B+D = bulge + disk, and A = ambiguous) and ($a$) optical
  spectral types among the ULIRGs from Paper I, ($b$) {\em IRAS}
  25-to-60 $\mu$m colors among the ULIRGs from Paper I, ($c$) {\em
    IRAS} 25-to-60 $\mu$m colors among PG~QSOs, ($d$) infrared excess,
  $L_{\rm IR}/L_{\rm B}$ among PG~QSOs, ($e$) FIR strength, $L$(60
  $\mu$m)/$L$(15 $\mu$m), among PG~QSOs, and ($f$) bolometric
  luminosity among PG~QSOs.  Radio-loud PG~QSOs are indicated by an
  ``R''. Panels ($a$) and ($b$) show that the hosts of warm,
  quasar-like ULIRGs all have a prominent spheroidal component, while
  the other panels indicate that QSOs with small infrared excesses,
  undetected FIR emission, and high bolometric luminosities favor
  elliptical hosts. Radio-loud QSOs avoid late-type systems with
  disks.}
\end{figure*}

\begin{figure*}[ht]
\epsscale{1.0}
\plotone{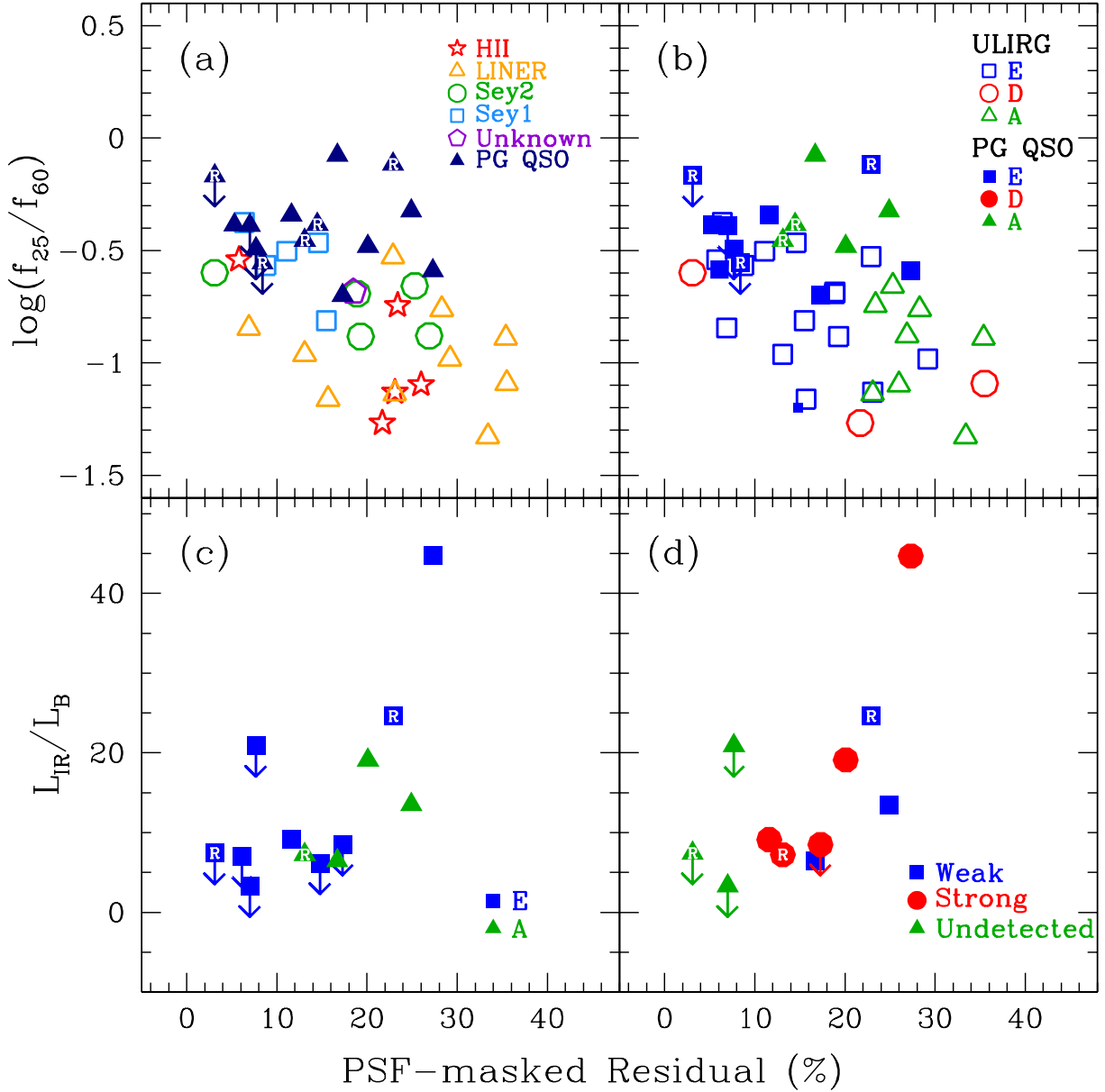}
\caption{ The PSF-masked residuals (as defined in the text) plotted
  against ($a$) and ($b$) the {\em IRAS} 25-to-60 $\mu$m colors and
  ($c$) and ($d$) the infrared excesses. The top panels contain all
  ULIRGs and PG~QSOs in our study while the bottom two panels show
  only the PG~QSOs. The data are labeled either by optical spectral
  type, morphological class, or FIR strength. Radio-loud PG~QSOs are
  indicated by an ``R''. The residuals for the 9 QSOs that are bulge +
  disk systems (B + D) are not shown in these panels since the
  residuals from the two galaxy component fits are necessarily smaller
  than those from the one galaxy component fits.  The residuals are
  smaller among warm, quasar-like ULIRGs and QSOs ($a$) with dominant
  elliptical morphology ($b$), small infrared excesses ($c$) and
  undetected FIR emission ($d$). }
\end{figure*}

\begin{figure*}[ht]
\epsscale{1.0}
\plotone{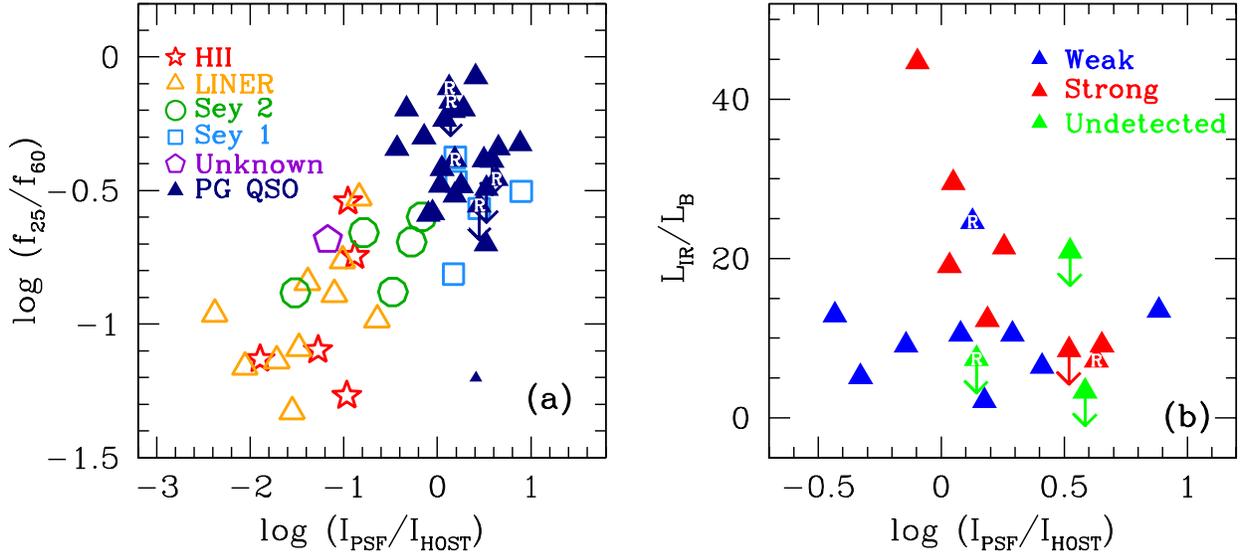}
\caption{ The intensity of the PSF component normalized to that of the
  host galaxy, $I_{PSF}/I_{\rm host}$, is plotted against ($a$) the
  {\em IRAS} 25-to-60 $\mu$m colors of ULIRGs and PG~QSOs and ($b$)
  the infrared excess, $L_{\rm IR}/L_{\rm B}$, of PG~QSOs only. The
  data are labeled either by optical spectral type ($a$) or FIR
  strength ($b$). Radio-loud PG~QSOs are indicated by an ``R''. Warm,
  quasar-like ULIRGs and PG~QSOs have stronger PSF components than
  H~II and LINER ULIRGs. Infrared-excess QSOs tend to have weaker PSF
  components than infrared-faint QSOs on average, although significant
  scatter is seen.}
\end{figure*}

\begin{figure*}[ht]
\epsscale{0.8}
\plotone{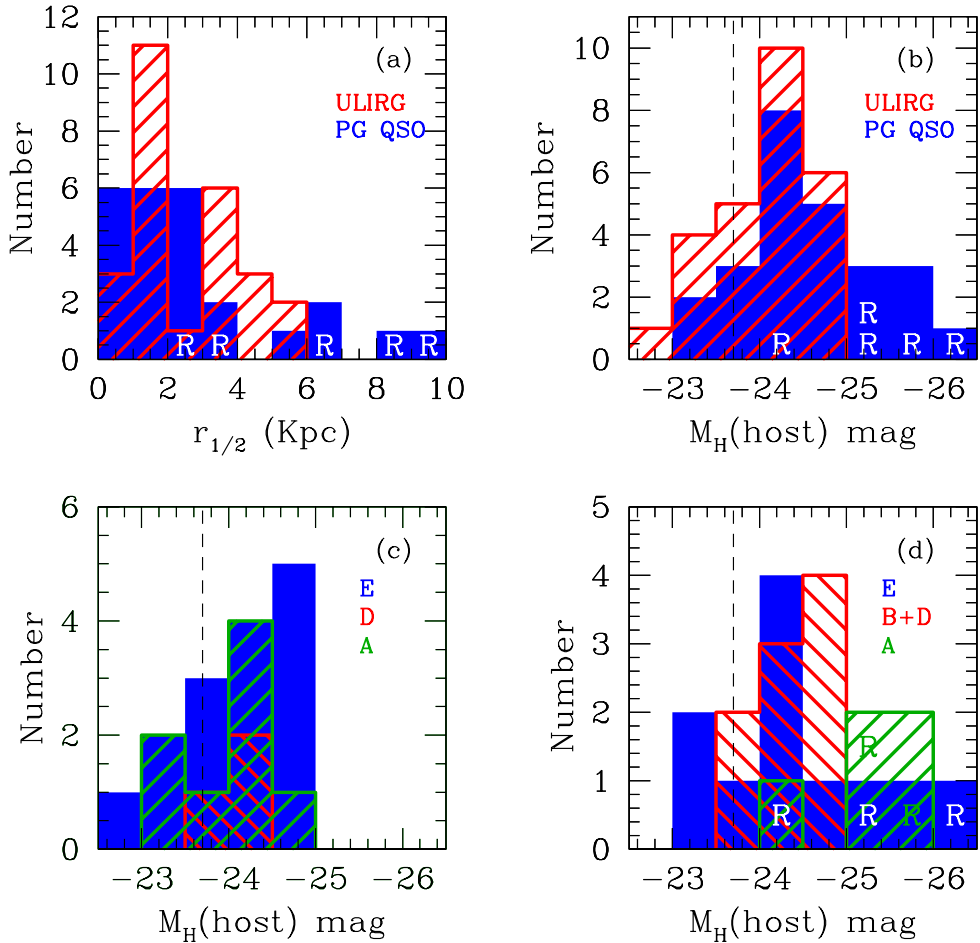}
\caption{ Distribution of the half-light radii (spheroid component
  only) and total host (spheroid + disk + tidal features, if relevant)
  absolute magnitudes for all ULIRGs and PG~QSOs in the {\em HST}
  sample.  Radio-loud PG~QSOs are indicated by an ``R''. The vertical
  dashed line in panels ($b$), ($c$), and ($d$) represents $M^*_H =
  -23.7$ mags, the H-band absolute magnitude of a $L^*$ galaxy in a
  Schechter function description of the local field galaxy luminosity
  function. In panels ($a$) and ($b$), the ULIRGs are cross-hatched
  red and the PG~QSOs are in blue. A K-S test on these data indicates
  no significant difference between the host sizes and magnitudes of
  the 1-Jy ULIRGs and radio-quiet PG~QSOs in this sample. The
  radio-loud QSOs are, however, significantly larger and brighter than
  the ULIRGs and radio-quiet QSOs. Panel ($c$) shows the distribution
  of host absolute magnitudes for ULIRGs according to their morphology
  (blue represents elliptical, red hatched corresponds to late type,
  and green hatched indicates ambiguous systems). Panel ($d$) is the
  same as panel ($c$) but for the PG~QSOs. No obvious trends with
  dominant morphological type are seen in the ULIRG data. QSOs with
  elliptical hosts show a broad range of host absolute magnitude,
  while QSOs with bulge + disk and ambiguous morphology tend to
  populate the lower and upper ends of the host luminosity
  distribution, respectively. }
\end{figure*}

\begin{figure*}[ht]
\epsscale{0.9}
\plotone{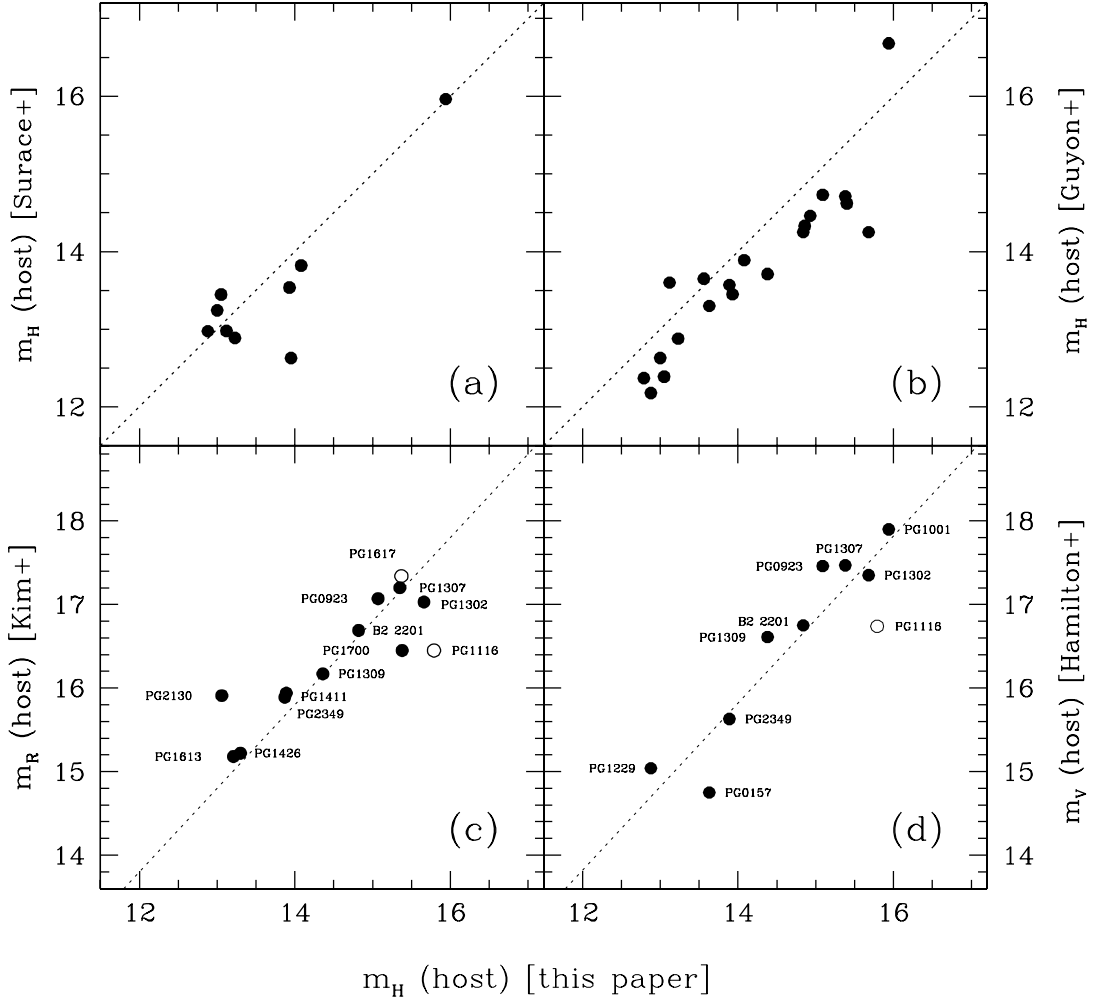}
\caption{ Comparison of the host magnitudes derived from our {\em HST}
  data with the results derived from ($a$) tip-tilt H-band imaging by
  Surace et al.\ (2001), ($b$) adaptive optics H-band imaging by Guyon
  et al.\ (2006), ($c$) archived {\em HST} R-band imaging of Kim et
  al.\ (2008), and ($d$) archived {\em HST} V-band imaging of Hamilton
  et al.\ (2008). An excellent match is found with the data of Surace
  et al., except for one object, PG~0007+106 (open circle in
  $a$). This is an optically violently variable source so the
  difference may be due to uncertainties in the removal of the central
  PSF in Surace et al.\ data. The Guyon et al.\ values are
  systematically $\sim$ 0.4 mag. brighter than our measurements. Given
  the good agreement between our data and those of Surace et al., we
  suspect that this shift is either due to uncertainties in the PSF
  subtraction from the AO data of Guyon et al. or systematic
  underestimate of their background level. The median R$-$H color
  (V$-$H) of the 13 (10) QSOs in $c$ ($d$) is 1.9 (1.9) mag.\ and is
  shown as the dashed diagonal line. The symbols in these two panels
  indicate the reliability of the H-band host magnitudes (filled
  circle = reliable, open circles = less reliable due to PSF mismatch;
  these data points were not used in the calculations of the
  averages).}

\end{figure*}
 
\begin{figure*}[ht]
\epsscale{0.75}
\caption{H-band FP of elliptical galaxies. In both panels, the large
  solid symbols are ULIRGs and PG~QSOs with ``pure'' elliptical ($n$ =
  4) hosts from the present NICMOS sample. The bulges of the bulge +
  disk systems are excluded to avoid uncertainties related to the
  bulge/disk decomposition. The open symbols are the optical mergers
  of Rothberg \& Joseph (2006; purple stars) and the
  optically/X-ray/radio more luminous QSOs from Dunlop et al.\ (2003;
  brown diamands), Hamilton et al.\ (2008; green squares), and Wolf \&
  Sheinis (2008; red triangles). The small dots represent the data
  from Pahre (1999; blue), Bernardi et al.\ (2003; orange), and
  Zibetti et al.\ (2002; cyan) on inactive spheroids. See text for
  assumed color transformation. Radio-loud PG~QSOs are indicated by an
  ``R''. The dotted line is a linear fit through the data on inactive
  spheroids, the solid line is a linear fit through the NICMOS data on
  the PG~QSOs, and the short-dashed line is a linear fit through the
  ULIRGs. The slope of the relation for ULIRGs is indistinguishable
  from that of the QSOs but is significantly steeper than for inactive
  spheroids.  A systematic shift at small half-light radii is also
  seen in panel ($b$), where the stellar velocity dispersions on the
  ULIRGs and PG~QSOs from Dasyra et al.\ (2006b, 2007) are compared
  with the FP of inactive spheroids. This trend is also seen among the
  optical mergers of Rothberg \& Joseph (2006). }

\end{figure*}

\begin{figure*}[ht]
\epsscale{1.0}
\plotone{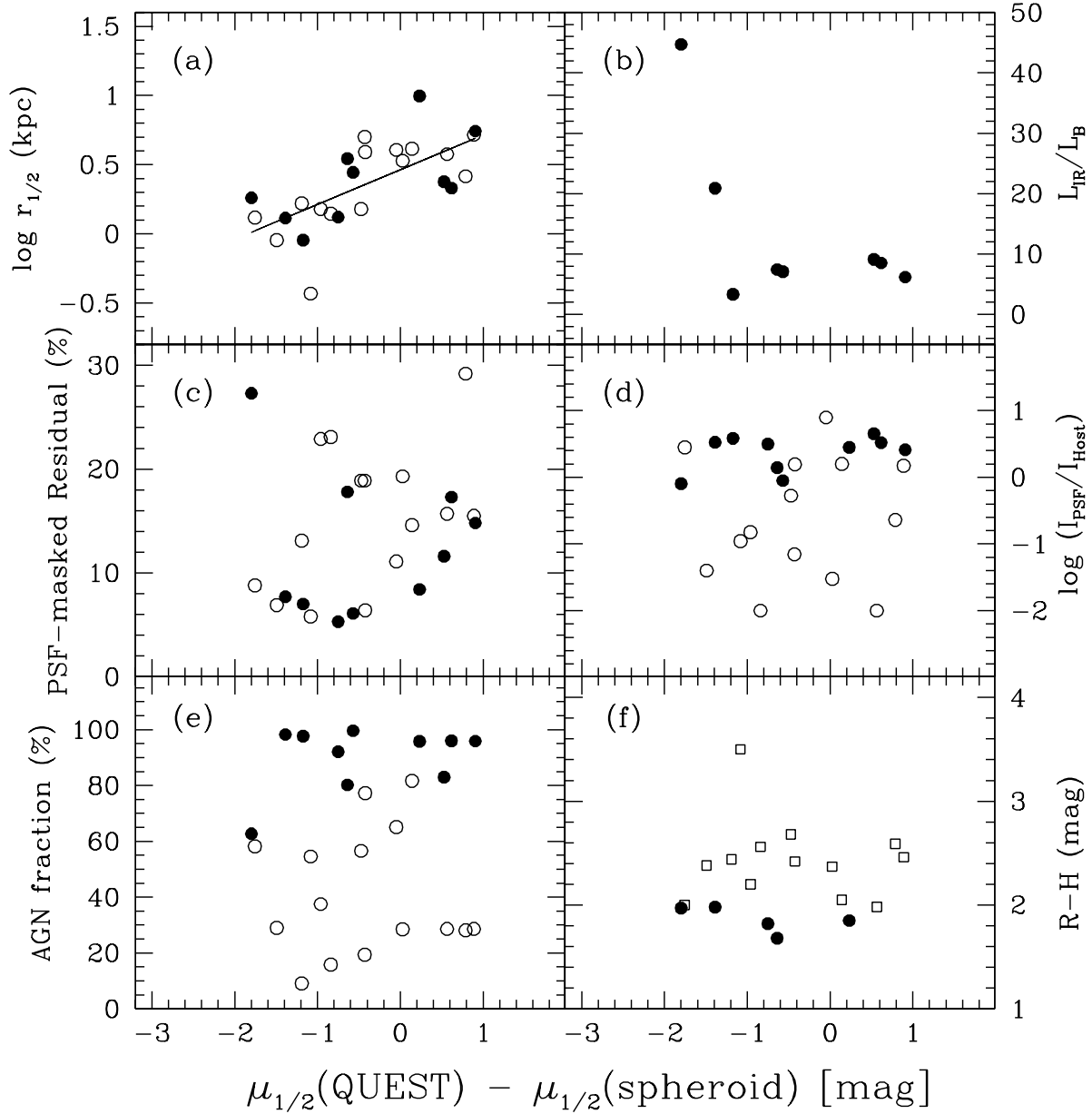}
\caption{ Surface brightness deviation as a function of ($a$)
  half-light radius, ($b$) infrared excess, ($c$) PSF-masked residual,
  ($d$) PSF-host flux ratio, ($e$) {\em Spitzer}-derived AGN
  fractional contribution to the bolometric luminosity (from Veilleux
  et al.\ 2009), and ($f$) R$-$H host colors for the ULIRGs from
  Veilleux et al.\ (2002; open squares), Veilleux et al.\ (2006; open
  circles), and PG~QSOs of this paper (filled circles). The surface
  brightness excess is defined as the difference between the observed
  surface brightness and the surface brightness expected of an
  inactive spheroidal galaxy with the same half-light radius, as
  determined by a linear regression through the data of Pahre (1999)
  or Bernardi et al.\ (2003) in Figure 8$a$. The bulges of the bulge +
  disk systems are excluded to avoid uncertainties related to the
  bulge/disk decomposition. An obvious trend is seen with galaxy size
  (the probability that this correlation is fortuitous $P[null]$ =
  0.02\%) and perhaps also with infrared excess though the statistics
  for this latter quantity are poor. }
\end{figure*}

\begin{figure*}[ht]
\epsscale{0.95}
\plotone{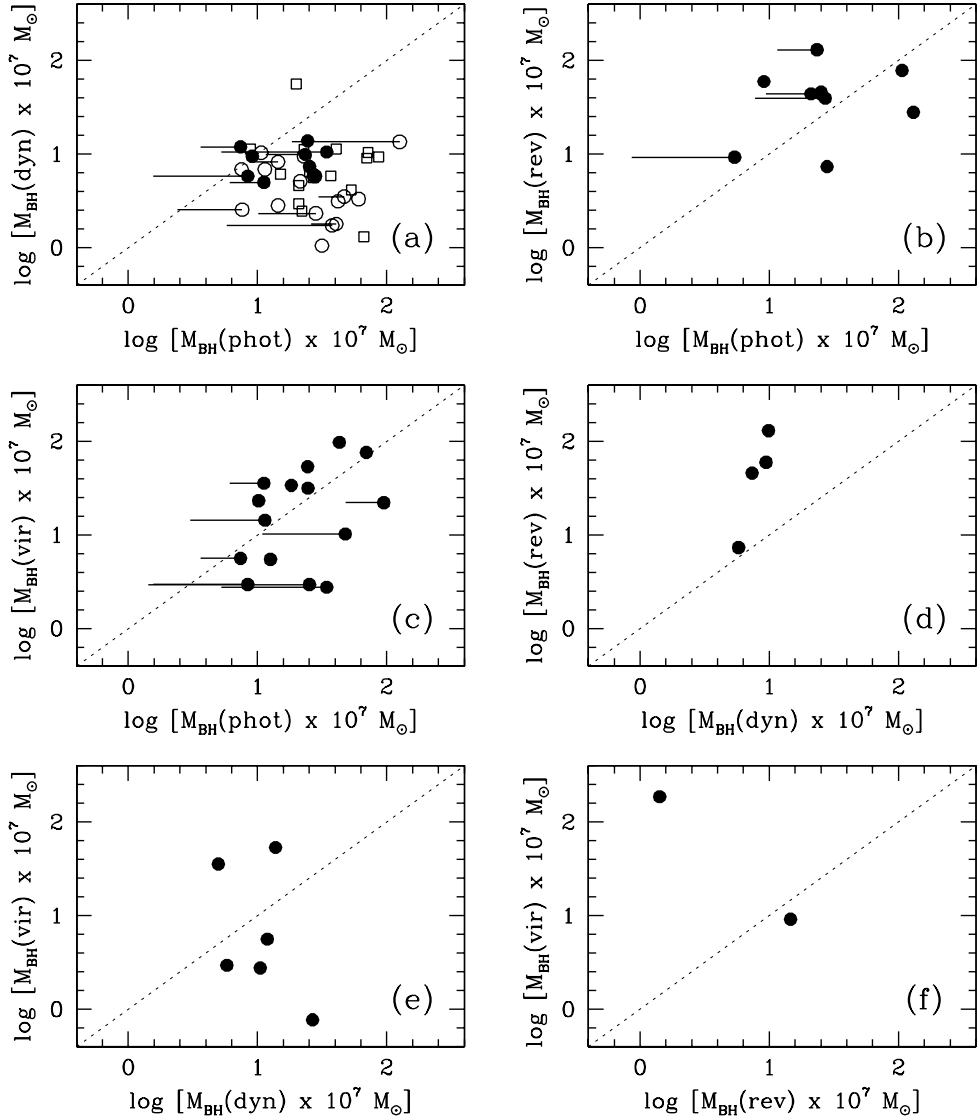}
\caption{ Comparison of black hole masses in ULIRGs and PG~QSOs
  derived from the photometric measurements of Paper I and the present
  paper, the stellar dynamical measurements of Dasyra et al.\ (2006a,
  2006b, 2007), the reverberation mapping measurements of Peterson et
  al.\ (2004) updated by Bentz et al.\ (2006), and the virial masses
  of Vestergaard \& Peterson (2006). ULIRGs from Veilleux et al.\
  (2002) are the open squares, ULIRGs from Veilleux et al.\ (2006) are
  the open circles, and PG~QSOs of this paper are the filled
  circles. Systems with either ``pure'' elliptical or bulge + disk
  hosts are included here. Discrepancies are seen in all panels,
  particularly those involving the dynamical measurements of large
  black hole masses. The best agreement (to within a factor of 3) is
  seen when comparing the photometric, reverberation-mapping, and
  virial mass estimates (panels $b$ and $c$). Horizontal segments on
  the photometric mass estimates indicate the effect of correcting the
  host magnitudes for possible excess H-band emission from a young
  circumnuclear stellar population (using the surface brightness
  deviations presented in Figure 9). }
\end{figure*}

\begin{figure*}[ht]
\epsscale{0.95}
\plotone{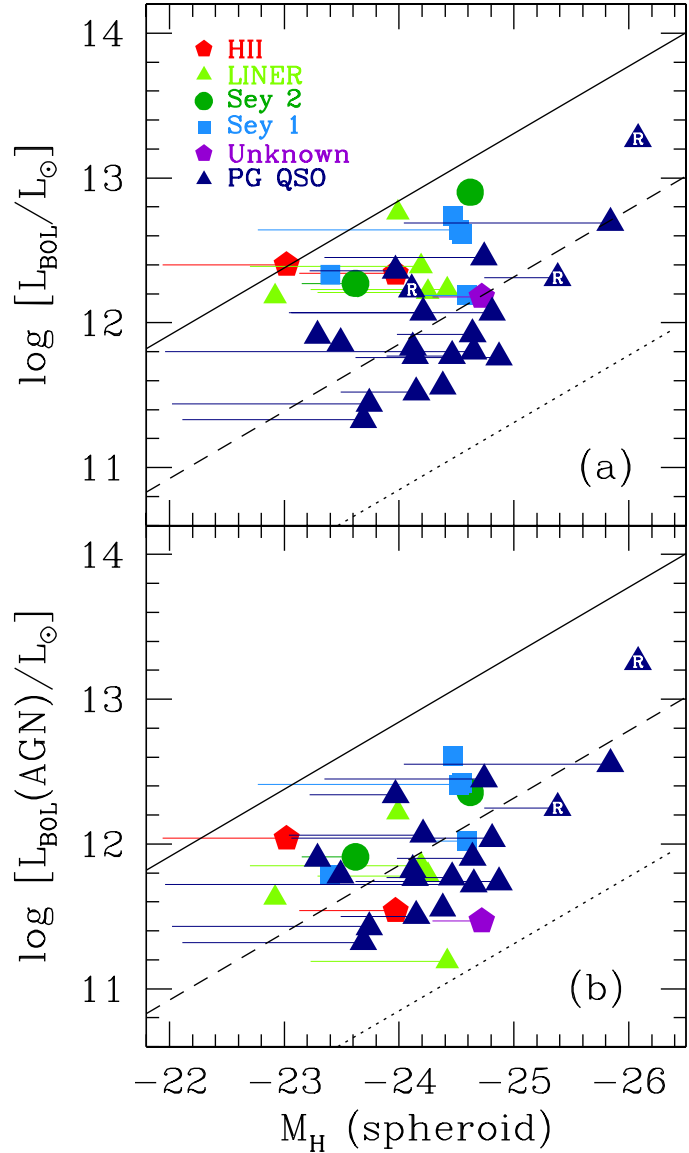}
\caption{ ($a$) Total bolometric luminosities and ($b$) AGN-only
  bolometric luminosities in ULIRGs and PG~QSOs with elliptical hosts
  {\em versus} absolute H-band magnitudes of the spheroidal components
  derived from the NICMOS data. Radio-loud PG~QSOs are indicated by an
  ``R''. Systems with either ``pure'' elliptical or bulge + disk hosts
  are included here. Diagonal dotted, dashed, and solid lines
  represent 1\%, 10\%, 100\% of the Eddington luminosity using the
  relation of Marconi \& Hunt (2003) to translate spheroid magnitudes
  into black hole masses.  The AGN fractional contributions to the
  bolometric luminosities used to produce panel ($b$) were taken from
  Veilleux et al.\ (2009). None of the objects in the sample radiate
  at super-Eddington rates. The Eddington ratios of the radio-quiet
  and radio-loud QSOs are statistically the same as those of ULIRGs on
  average (of order 3-30\%). Horizontal segments on the spheroid
  magnitudes indicate the effect of correcting for possible excess
  H-band emission from a young circumnuclear stellar population (using
  the surface brightness deviations presented in Figure 9). }
\end{figure*}

\end{document}